\documentclass[%
 aip,
 jmp,%
 amsmath,amssymb,
 reprint,%
]{revtex4-1}
\usepackage[version=3]{mhchem} 
\usepackage{mciteplus} 
\usepackage{verbatim} 
\usepackage{color} 
\usepackage[dvipsnames]{xcolor}
\usepackage{hyperref}
\usepackage{bm}
\usepackage{multirow}
\usepackage{booktabs}
\usepackage{graphicx}
\newcommand{\citen}[1]{\hspace{-1 ex} \citenum{#1}} 
\mciteSetSublistMode{s}
\newcommand{\trm}{\textrm}

\newcommand{\cisPt}{$\trm{Pt}(\trm{NH}_3)_2\trm{Cl}_2$}
\newcommand{\aquaPt}{$[\trm{Pt}(\trm{NH}_3)_2(\trm{OH}_2)\trm{Cl}]^+$}

\newcommand{\monoPt}{$[\trm{Pt}(\trm{NH}_3)_2(\trm{N-heterocycle})\trm{Cl}]^+$}
\newcommand{\aminoPt}{$[\trm{Pt}(\trm{NH}_3)_3\trm{Cl}]^+$}
\newcommand{\pyriPt}{$\trm{\textit{cis}-}[\trm{Pt}(\trm{NH}_3)_2(\trm{pyridine})\trm{Cl}]^+$}

\newcommand{\gPtO}{$g_\trm{PtO}(r)$}
\newcommand{\gPtH}{$g_\trm{PtH}(r)$}
\newcommand{\nPtO}{$n_\trm{PtO}$}
\newcommand{\nPtH}{$n_\trm{PtH}$}

\newcommand{\nOOw}{$n_{\trm{OO}_\trm{w}}$}
\newcommand{\nOHw}{$n_{\trm{OH}_\trm{w}}$}
\newcommand{\nHOw}{$n_{\trm{HO}_\trm{w}}$}
\newcommand{\nClHw}{$n_{\trm{ClH}_\trm{w}}$}

\newcommand{\sh}{\textit{s}}
\newcommand{\mcol}{\multicolumn}
\newcommand{\rint}{r_\trm{int}}
\newcommand{\rmax}{r_{\trm{max}}}
\newcommand{\kB}{k_\trm{B}}
\begin{document}
\title{Blue moon ensemble simulation of aquation free energy profiles 
applied to mono and bifunctional platinum anticancer drugs}
\author{Teruo Hirakawa}
\affiliation{Department of Precision Science and Technology, 
Graduate School of Engineering, Osaka University, 2-1, 
Yamada-oka, Suita, Osaka 565-0871, Japan}
\affiliation{Institut des Sciences Mol\'{e}culaires (ISM), 
Universit\'{e} Bordeaux, CNRS UMR 5255, 351 cours de la Lib\'{e}ration, 
33405 Talence cedex, France}
\author{David R. Bowler}
\affiliation{Department of Physics \& Astronomy,  University College London (UCL),
Gower St, London, WC1E 6BT, UK}
\affiliation{London Centre for Nanotechnology, UCL, 17-19 Gordon St,
 London WC1H 0AH, UK}
\affiliation{Centre for Materials Nanoarchitechtonics (MANA), National Institute for Materials Science 
(NIMS), 1-1 Namiki, Tsukuba, Ibaraki 305-0044, Japan}
\author{Tsuyoshi Miyazaki}
\affiliation{Computational Materials Science Unit (CMSU), NIMS, 1-1 Namiki, Tsukuba, Ibaraki 305-0044, Japan}
\affiliation{Centre for Materials Nanoarchitechtonics (MANA), National Institute for Materials Science 
(NIMS), 1-1 Namiki, Tsukuba, Ibaraki 305-0044, Japan}
\author{Yoshitada Morikawa}
\affiliation{Department of Precision Science and Technology, 
Graduate School of Engineering, Osaka University, 2-1, 
Yamada-oka, Suita, Osaka 565-0871, Japan}
\affiliation{Elements Strategy Initiative for Catalysts and Batteries (ESICB), 
Kyoto University, Katsura, Kyoto 615-8520, Japan}
\affiliation{Research Center for Ultra-Precision Science and Technology, 
Graduate School of Engineering, Osaka University, 2-1, 
Yamada-oka, Suita, Osaka 565-0871, Japan}
\author{Lionel A. Truflandier}
\email{lionel.truflandier@u-bordeaux.fr}
\affiliation{Institut des Sciences Mol\'{e}culaires (ISM), 
Universit\'{e} Bordeaux, CNRS UMR 5255, 351 cours de la Lib\'{e}ration, 
33405 Talence cedex, France}
\affiliation{Department of Precision Science and Technology, 
Graduate School of Engineering, Osaka University, 2-1, 
Yamada-oka, Suita, Osaka 565-0871, Japan}
%
%
%
%
\date{\today}
\begin{abstract}
Aquation free energy profiles of neutral cisplatin and cationic monofunctional derivatives,
including triaminochloroplatinum(II) and cis-diammine(pyridine)chloroplatinum(II),
were computed using state of the art thermodynamic integration, for which temperature and 
solvent were accounted for explicitly using density functional theory based canonical 
molecular dynamics (DFT-MD). For all the systems the "inverse-hydration" where the 
metal center acts as an acceptor of hydrogen bond has been observed. This has 
motivated to consider the inversely bonded solvent molecule in the definition of the
reaction coordinate required to initiate the constrained DFT-MD trajectories.
We found that there exists little difference in free enthalpies of activations,
such that these platinum-based anticancer drugs are likely to behave the same way in 
aqueous media. Detailed analysis of the microsolvation structure of the square-planar complexes, 
along with the key steps of the aquation mechanism are discussed.
\end{abstract}
\maketitle
%

\section{Introduction}
\begin{figure}[tbh!]
	\centering
	\includegraphics[width=0.80\textwidth]{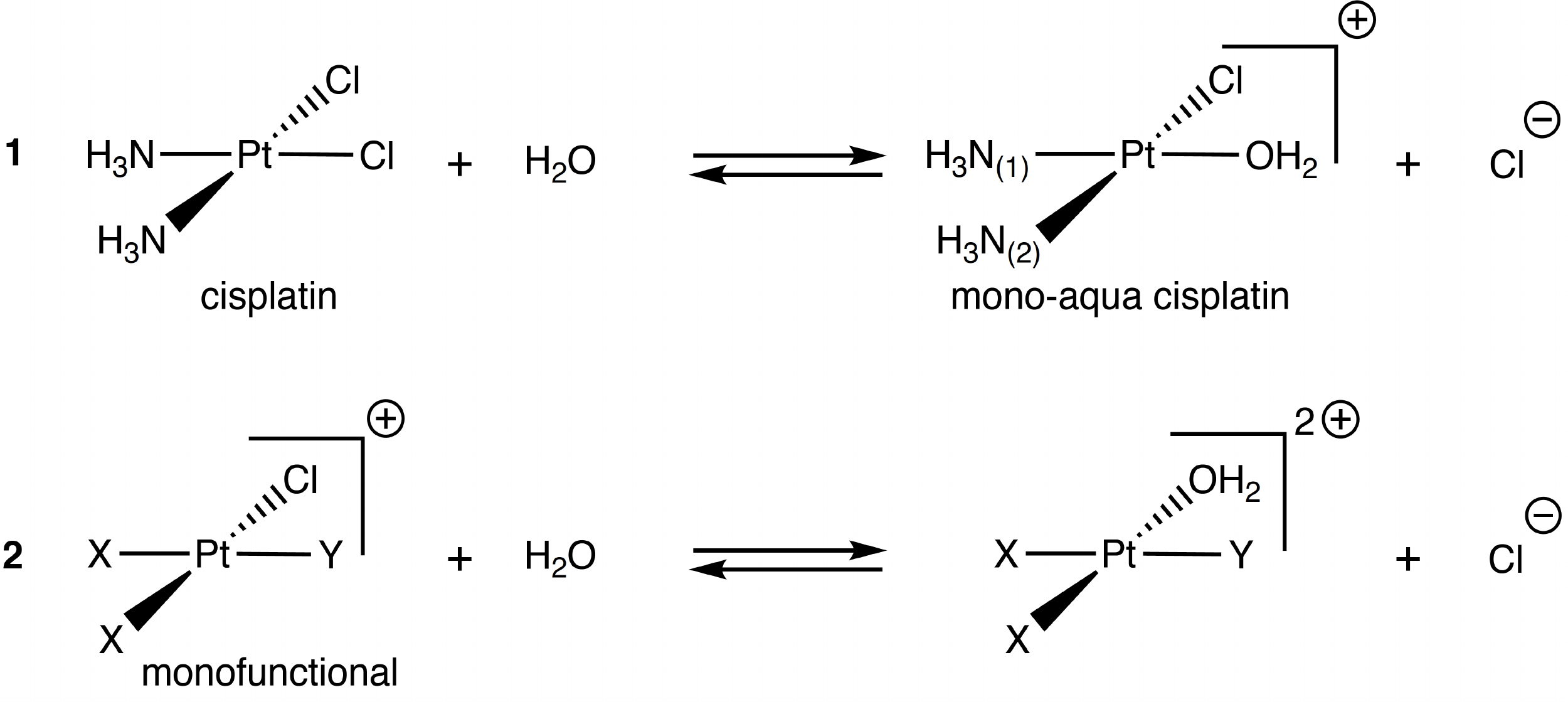}
	\caption{Aquation reaction of cisplatin (\textbf{1}) and monofunctional derivatives (\textbf{2}).} 
	\label{fig:scheme1}  
\end{figure}

The discovery of the anticancer activity of cisplatin (Scheme~\ref{fig:scheme1}) 
during the 60's has promoted fast development of platinum(II)-based drugs which are currently found 
in chemotherapy regimens 
(see Refs. \citen{wheate_status_2010,johnstone_anticancer_2014,apps_Ptdrugs_2015,johnstone_chemrev_2016} 
for reviews). Despite few successes for some of them in curing specific cancers, 
eg. the treatment of testicular cancer with cisplatin, their efficiency against the 
broad spectrum of carcinoma remains limited. The main challenges to overcome are: (1) the elimination 
of the severe side effects related to their poor selectivity with respect to the tumor cells, and (2) the 
resistance ---either intrinsic (static) or evolutionary (dynamic)---, observed for some types of 
cancer.\cite{brabec_resistance_2005,chabner_nature_2005,kelland_nature_2007} 
Solving the first problem requires the design of new drug delivery strategies,\cite{apps_Ptdrugs_2015,johnstone_chemrev_2016} 
which ideally prevents the degradation of platinum complex in the blood stream and allows 
for a precise targeting of the tumor tissue. Solution to the second problem 
implies heuristic methods, such as the structure-activity relationship 
(SAR),\cite{hambley_coordchemrev_1997,johnstone_anticancer_2014} from which
a huge number\cite{hambley_coordchemrev_1997} of platinum(II)-based drugs and 
platinum(IV)-based prodrugs were synthesized and tested as many attempts to mimic the 
cisplatin's mechanism of action while trying to improve the cytotoxic properties. 
As a result of 30 years of trial and error less than 30 platinum drugs have been considered for clinical trials, 
with only 2 of them (carboplatin and oxaliplatin) 
approved wordwilde for clinical used.\cite{wheate_status_2010,johnstone_chemrev_2016} Whereas high-throughput 
synthesis and screening\cite{ziegler_jbic_2000} of drug candidates may provide rapid ---but partial--- solutions to an urgent
problem, understanding and controlling every details of cisplatin's mechanism constitutes a safer ---but far more longer--- 
route towards a rational design of a universal platinum-based anticancer molecule.
The process by which cisplatin (and bifunctional derivatives) leads to cell death is now rather well 
understood.\cite{jamieson_chemrev_1999,dhar_review_2009} 
It is divided into 4 main steps: (i) the cellular uptake, (ii) the aquation/activation (Scheme~\ref{fig:scheme1}) 
in the cellular media, (iii) the DNA platination, that is, bifunctional intra- and interstrand crossed-links $\textit{via}$ the 
fomation of 2 covalent bonds between the metal center and the purine bases, and (iv) the DNA-damage 
recognition initiating apoptosis, or cell-cycle arrest eventually followed by an attempt to repair the lesion. 
In elucidating each step of the mechanism, atomistic computations and simulations 
based on either (classical) molecular, quantum, or mixed molecular/quantum mechanics 
can provide important insights.

At the pure quantum mechanics level of theory, some of the efforts were devoted to investigate the hydration
structure of the square-planar Pt-complexes\cite{vidossich_accchemressaccounts_2016,*beret_jctc_2008,*beret_cpc_2009,
*vidossich_chemphyschem_2011,truflandier_inorgchem_2011,sutter_chemphyschem_2011} in response to some 
experimental evidence\cite{baidina_crystal_1981,rizzato_angew_2010} showing unusual bonding situation ---referred 
as to inverse-hydration--- where the Pt atom acts as a hydrogen-bond acceptor. 
Energetic contributions\cite{kozelka_angew_2000,lopes_theoretical_2008,aono_jctc_2016} 
and topological analysis,\cite{berges_inorgchem_2013} along with minute details on the influence 
of the bulk solvation effects have been extensively discussed in literature.\cite{melchior_jctc_2015,
kroutil_inorgchem_2016,aono_jctc_2016}
Another interest which is also closely related to our concern is the activation of cisplatin after
the cellular uptake of the platinum drug. The aquation reaction (also loosely called hydrolysis) represented 
on Scheme~\ref{fig:scheme1} through the substitution of one or both chloride ligands of cisplatin by water 
molecules was recognized as a crucial step to initiate DNA platination.\cite{bancroft_jacs_1990} 
The proof of concept is carboplatin\cite{knox_mechanism_1986,alberts_new_1998,johnstone_oxidative_2014} 
featuring the cyclobutanedicarboxylate in replacement of the chloride leaving ligands of cisplatin.
It demonstrates much more slower aquation and DNA platination rates 
but displayed similar crossed-link sites.\cite{blommaert_jacs_1995} 

To access more details on the aquation of cisplatin and derivatives, many theoretical works have focused 
on computing activation barriers using static molecular clusters and implicit solvation 
models.\cite{zhang_hydrolysis_2001,
  *robertazzi_hydrogen_2004,         *lau_hydrolysis_2006,
  *alberto_second-generation_2009,*lucas_neutral_2009,
  *banerjee_cpl_2010,*melchior_comparative_2010} 
When applied for modelling aqueous reactions, insights arising from this type 
of computations remain limited since they do not account properly for the dynamic 
reorganisation of solvation shell surrounding the molecule, especially when the solvent
is one of the reactant.
More advanced but computationally demanding methods can be envisaged where
bulk and microsolvation effect are treated on equal footing using density functional 
theory molecular dynamics (DFT-MD) along with free energy path
evaluation based on metadynamics further refined by umbrella 
sampling.\cite{carloni_key_2000,lau_pccp_2010} Alternatives to DFT-MD, 
such as the reference interaction site model self-consistent 
field\cite{yokogawa_jcp_2007,yokogawa_jcp_2009} (RISM-SCF)
can also be employed.\cite{yokogawa_dalton_2011}

\begin{figure}
	\centerline{\includegraphics[width=0.90\textwidth]{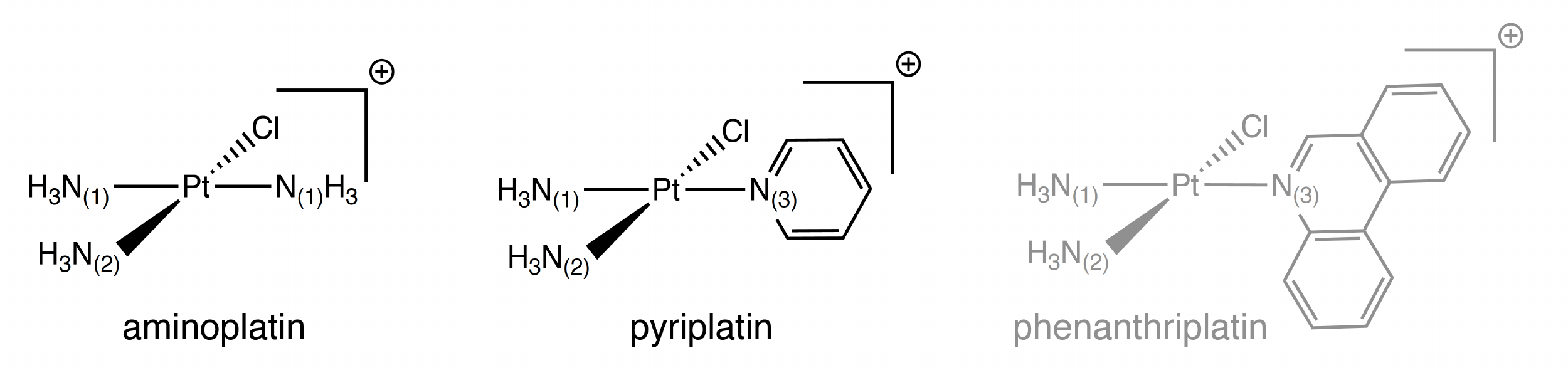}}
	\caption{Structures cationic monofunctional platinum complexes. Note that phenantriplatin
	has not been investigated in this work.} 
	 \label{fig:scheme2}  
\end{figure}

From the last ten-years research focussing on reducing the tumor-cell resistance to 
bifunctional cisplatin and derivatives, it has been shown that monofunctional 
analogues\cite{hollis_jmedchem_1989,brabec_dna_2002,bursova_biophys_2005} 
can be considered as potent candidates.\cite{lovejoy_pnas_2008,wang_pnas_2010,zhu_cancer_2012} 
Monofunctional platinum(II)-based drugs are cationic square-planar complexes (Scheme.~\ref{fig:scheme2})
of formula \monoPt, deriving from the triammine-chloro precursor \aminoPt, later referred
as to aminoplatin. In a systematic investigation of the cancer-cell responses with respect to 
the heterocyclic N-donor ligand bound to the metal, it has been demonstrated\cite{park_pnas_2012}
that the size and the arrangement of the aromatic rings affects drastically 
the cytotoxic and selectivity properties of the monofunctional platinum(II) drugs.
As a matter of fact, whereas aminoplatin was found to be biologically inactive,
phenanthriplatin, and to a lesser extent pyriplatin, presents a remarkable potency,\cite{johnstone_jacs_2014}
exceeding in the majority of the cases the anticancer activity of cisplatin.\cite{park_pnas_2012,johnstone_jacs_2014}

This work aims to investigate the aquation reaction of Pt(II) complexes using state of the art thermodynamic 
integration coupled to DFT-MD simulations to determine if cationic monofunctional derivatives present differences 
in the free energy profiles and mechanism of reaction. The paper is organized as following: Section 2 provides 
methodological and computational aspects related to the DFT-BOMD simulation and blue moon ensemble (BME) 
integration. Results and discussion are presented in Section 3. The paper concludes with a brief summary
of the key findings and an outlook.
\section{Computational details}
\subsection{DFT-BOMD simulations}
\label{sec:bomd-method}
All the DFT-MD simulations reported in this article were performed on the Born-Oppenheimer (BO)
surface, where the Kohn-Sham (KS) self-consistent-field (SCF) equations are solved at each 
step of the dynamics using the \textsc{Conquest} code.\cite{bowler_pssb_2006,bowler_jpcms_2010,hernandez_prb_1996} 
A strictly localized (atomic-like and finite range) numerical double-$\zeta$ basis 
set\cite{sankey_pao_1989,junquera_pao_2001,torralba_pao_2008} including 
polarisation functions (DZP) for expanding the valence wavefunctions along with norm-conserving 
pseudopotentials\cite{troullier_ncpp_1991} (NCPP) were especially designed for this work. 
Thereafter, this basis set will be referred as pseudo-atomic orbitals (PAO). Using a benchmark of 
isolated Pt complexes, reliability of the PAOs were checked against molecular calculations based 
on gaussian-type orbitals (GTO). Additional computational 
parameters related to the PAOs and NCPPs generation along with their assessments
can be found in the Sec. S1 of the Supporting Information (SI).
The KS-SCF equations were solved within the framework of the generalized gradient approximation (GGA) 
of the exchange-correlation functional proposed by Perdew, Burke, and Ernzerhof (PBE).\cite{perdew_generalized_1996}
This choice of GGA functional was more pragmatic than idealistic. 
It is well established (see Refs. \citen{lin_JCTC_2012,distasio_JCP_2014} for recent studies) that the PBE 
functional yields to a "glassy" state of liquid water at ambient temperature preventing a fully quantitative 
agreement with experiments. Nevertheless, also important
for this work, it has been demonstrated that the same functional is able to reproduce
the inverse-hydration feature.\cite{truflandier_probing_2010} 

Solvation of the platinum complexes has been modeled using cubic simulation boxes with
63 water molecules adapted in size to reach an average density of 1.0 g$\cdot$cm$^{-3}$. 
For cationic complexes, the excess of positive charge was balanced by 
substituting water molecule with hydroxide anion keeping the overall charge of the supercell neutral.
For the DFT-BOMD simulations, ionic cores were propagated using the velocity Verlet algorithm\cite{verlet_computer_1967} 
with a time step of 0.5 fs, for which a SCF convergence criteria of 10$^{-7}$ on the residual
of the electronic density, and a grid spacing of about 0.20 au were found sufficient to prevent 
any energy drift during microcanonical simulations.
The same set of convergence parameters were used for \textit{NVT}-ensemble simulations 
using a Nos{\'e}--Hoover chain\cite{martyna_nosehoover_1992,hirakawa_nosehoover_2017} 
of 5 thermostats with a frequency of 500 cm$^{-1}$.

\subsection{Blue moon ensemble integration}
\label{sec:bme-method}

The hydrolysis free energy profiles of cisplatin and its derivatives were 
calculated using thermodynamic integration \textit{via} the blue moon 
ensemble (BME) technique.\cite{carter1989constrained} Given a reaction 
coordinate, $\xi\equiv\xi(\{{\bf r}_{i}\})$, which in general depends 
on a subset of the $N$ atomic positions $\{\bf r_{\it i}\}$, the variation 
of free energy between some initial state, $\xi_0$, up to the current value 
of $\xi$ can be expressed as: 
\begin{equation}
  \Delta F (\xi,T)
  = \int^{\xi}_{\xi_0} 
 \left(\frac{\partial {F(\xi;T)}}{\partial{\xi}}\right)_{\xi'} d\xi',
 \label{eq:thermo}
\end{equation}
Integration is performed numerically through the sampling of the reaction coordinate 
using a discret set of $n_{\xi'}$ target values $\{\xi'\}$. The free energy gradient 
---also referred to as (minus) the mean force--- in Eq.~(\ref{eq:thermo}) is evaluated
for each $\xi'$ from a constrained MD trajectory where $\xi-\xi'=0$ is enforced.
Following BME Lagrangian formulation of Sprik and Ciccotti,\cite{sprik1998free} 
the mean force writes
\begin{equation}
\left( \frac{\partial {F(\xi;T)}}{\partial{\xi}}\right)_{\xi'} 
 =  \frac{ \langle Z^{-1/2} \big(-\lambda_{\xi}+ \kB T G \big ) 
 \rangle_{\xi'}}{{\langle Z^{-1/2} \rangle}_{\xi'}},
  \label{eq:bme}
\end{equation}
for which $\kB$ and $T$ are the Boltzmann constant and temperature, respectively.
The Lagrange multiplier, $\lambda_{\xi}$, associated with constrained coordinate $\xi$, 
defines the strength the constraint force updated at each MD step.\cite{ryckaert1977numerical,andersen1983rattle,ciccotti1986molecular} 
The remaining terms are given by,
\begin{eqnarray}
 Z & = & \sum^N_{i =1} \frac{1}{m_i}\left(\frac{\partial \xi}{\partial {\bf r}_{i}}\right)^2,\\
 G & = &\frac{1}{Z^2}\sum^N_{i = 1}\sum^N_{j = 1}\frac{1}{{m_i}{m_j}}
 \frac{\partial \xi}{\partial {\bf r}_{i}}\cdot
 \frac{{\partial}^2 \xi}{\partial {\bf r}_{i}{\bf r}_{j}}\cdot
 \frac{\partial \xi}{\partial {\bf r}_{j}}.
\end{eqnarray}
where $m_i$ designates the mass of the $i$th nucleus.  In Eq.~(\ref{eq:thermo}), the notation 
$\langle\cdot\rangle_{\xi'}$ stands for the canonical ensemble average performed for each target 
value $\xi'$. Around 20 values were used for the reaction path discretization. Ensemble averages performed 
at $T=310$ K were obtained from production runs of 2.5 ps subsequent to 1.0 ps of equilibration. 
Free energy errors were estimated by the method of Jacucci and Rahman.\cite{jacucci_comparing_1984}

\section{Results and discussion}

\subsection{MD analysis and reliability}
\label{sec:bomd-analysis}

\begin{table*}[htb!]
\scriptsize{
\caption{Comparison of selected structural parameters of the solvated bi and monofunctional cisplatin derivatives obtained 
with the PBE functional using various methods. For BOMD simulations, mean distance and angle values are given 
along with their respective standard deviations. Inequivalent nitrogen atoms are numbered according to Schemes \ref{fig:scheme1}
and \ref{fig:scheme2}.}
\label{table:struct_complexes}
	\vspace{12pt}
	\centerline{
	\begin{tabular}{lllccccc}
	\hline\hline
	Complex    &Method & Basis &\multicolumn{2}{c}{Distance (\AA)}& \multicolumn{2}{c}{Angle (deg)}
	 \\\hline
	cisplatin    &                    &                 &  Pt--N                &  Pt--Cl          & N--Pt--N     & Cl--Pt--Cl      \\
	       & BOMD-\textit{NVE} & PAO-DZP & $2.08\pm0.04$  & $2.37\pm0.03$ & $91.0\pm4.0$ & $92.5\pm3.4$ \\ 
	               & CPMD-\textit{NVE}$^a$ & PW 
	                &  $2.09\pm0.04$  & $2.35\pm0.03$ & $90.4\pm4.1$ & $92.3\pm4.3$  \\ \cmidrule{2-7}           
                    & Static-PCM$^b$ & GTO-DZP      &  $2.07$ & $2.34$ & $92.7$ & $94.7$ \\\cmidrule{4-7}
                    & Static        & GTO-DZP      &  $2.09$ & $2.30$ & $99.1$ & $95.7$  \\
                    & Static        & PAO-DZP     &  $2.10$ & $2.34$ & $98.5$ & $96.0$ \\\hline
mono-aqua cisplatin &    &     &  Pt--N$_{(1)}$[Pt--N$_{(2)}$]  &  Pt--Cl          & Pt--O & N--Pt--N   \\
%
  & BOMD-\textit{NVE} & PAO-DZP   & $2.05[2.08]\pm0.03[0.05]$  & $2.36\pm0.03$ & $2.10\pm0.07$ & $91.4\pm4.5$ \\ \cmidrule{2-7}    
                    & Static-PCM & GTO-DZP      &  $2.02[2.07]$ & $2.32$ & $2.10$ & $93.5$  \\ \cmidrule{4-7}
                    & Static        & GTO-DZP      &  $2.03[2.11]$ & $2.28$ & $2.10$ & $97.2$  \\
                    & Static        & PAO-DZP     &  $2.04[2.12]$ & $2.32$ & $2.11$ & $96.6$ \\\hline
aminoplatin   &  &  &  Pt--N$_{(1)}$[Pt--N$_{(2)}$]  &  Pt--Cl  & N$_{(1)}$--Pt--N$_{(2)}$ &  N$_{(1)}$--Pt--Cl   \\
%
   & BOMD-\textit{NVE} & PAO-DZP    & $2.06[2.08]\pm0.04[0.04]$  & $2.40\pm0.03$ & $90.5\pm4.2$ &  $89.5\pm4.2$ \\     \cmidrule{2-7}
                    & Static-PCM & GTO-DZP      & $2.06[2.08]$ & $2.33$ & $91.2$ & $88.8$ \\\cmidrule{4-7}
                    & Static        & GTO-DZP      &  $2.07[2.12]$ & $2.28$ & $93.6$ & $86.4$ \\
                    & Static        & PAO-DZP      &  $2.08[2.13]$ & $2.32$ & $93.4$ & $86.3$ \\\hline
pyriplatin       &  &  &  Pt--N$_{(1)}$[Pt--N$_{(2)}$]  &  Pt--Cl    & Pt--N$_{(3)}$ 
                    & N$_{(1)}$--Pt--N$_{(2)}$   \\
%
  & BOMD-\textit{NVE} & PAO-DZP     & $2.04[2.07]\pm0.03[0.04]$ & $2.37\pm0.03$ & $2.07\pm0.03$ & $90.3\pm4.1$ \\  \cmidrule{2-7}
                    & Static-PCM & GTO-DZP       & $2.07[2.08]$ & $2.33$ & $2.02$ & $91.3$ \\\cmidrule{4-7}
                    & Static        & GTO-DZP       & $2.08[2.10]$ & $2.28$ & $2.02$ & $95.0$  \\
                    & Static        & PAO-DZP       & $2.09[2.13]$ & $2.32$ & $2.02$ & $95.0$ \\\hline
\hline\hline
\end{tabular}                  
}
}
$^a$Results obtained in Ref.~\citen{truflandier_inorgchem_2011} using the Car-Parrinello (CP)
approach as implemented in \textsc{Quantum ESPRESSO}\cite{qe_short_2009}. The implementation 
is based on planewave (PW) basis set and pseudopotentials. The same model has been used for the BOMD 
and CPMD simulations, eg. box size and number of atoms. $^b$'Static' refers to gas-phase optimized 
structures including (or not) an implicit water-solvent treatment based on the Polarizable Continuum 
Model (PCM) as implemented in \textsc{GAUSSIAN09}\cite{gaussian09_short}.
\end{table*}
To assess the quality of our DFT-BOMD simulations, the liquid structure of the solvated Pt-complexes
were analysed \textit{via} the radial distribution function (RDF) ---leading to the pair correlation 
function $g(r)$--- and the power spectral density (PSD) of the velocity autocorrelation 
function (VACF) ---leading to the vibrational density of states (VDOS). The methodology 
used for the calculations of these properties can be found in the SI of Ref.~\citen{truflandier_probing_2010}. 
For each system, RDFs (VDOS) were computed from a \textit{NVT} (\textit{NVE}) 
production run of 10 ps performed after a \textit{NVT} equilibration step of 15 ps.
The PSD was refined using the Welch window function with 8 non-overlapping segments 
of 512 data points.\cite{press_numerical_1992} Analysis and a brief discussion of the 
solvent RDFs and VDOS in light of previous studies are given in Sec. S2 of the SI. 
Below we shall concentrate on the solute structural and vibrational signatures. 

\begin{figure}[htb!]
	\centerline{\includegraphics[width=0.90\textwidth]{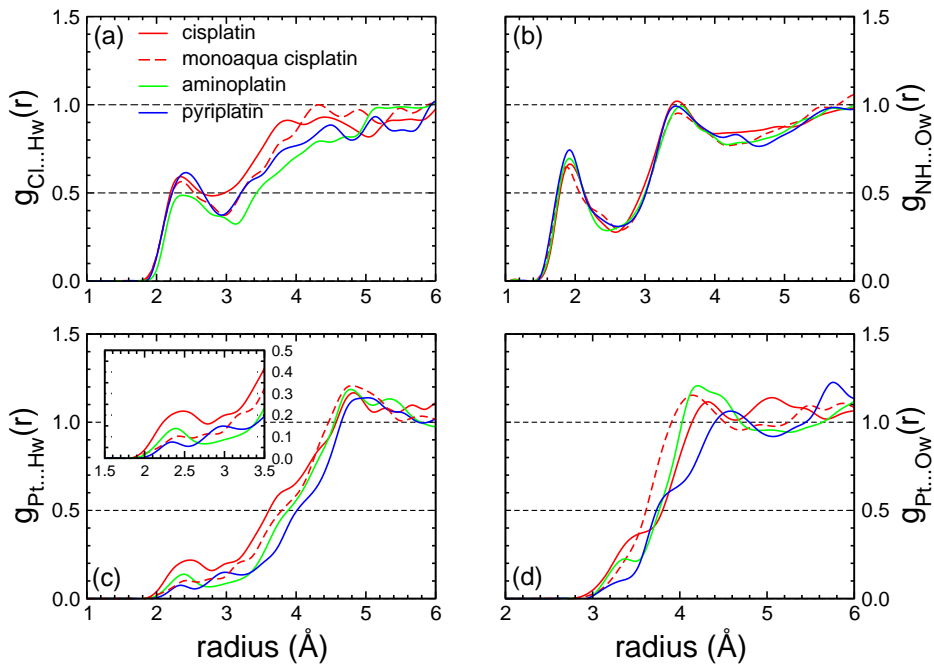}}
	\caption{Selected $g(r)$ describing the solute-solvent dynamical structures for the 
	solvated bi and monofunctional cisplatin derivatives complexes studied in this work.} 
	 \label{fig:figure1}  
\end{figure}

Selected mean bond distances and angles obtained from the microcanonical BOMD simulations 
are collected in Table \ref{table:struct_complexes} for all the Pt-based reactants investigated in this work.
For cisplatin, when compared to previous DFT-MD simulations using the same XC functional but 
a different implementation,\cite{truflandier_inorgchem_2011} a nice agreement is observed.
By comparing across the set of platinum complexes the deviations of the Pt-ligand mean bond 
distances and angles with respect to gas phase optimized strutures, we can isolate two systematic 
effects of aqueous solvent: ($i$) an increase of the Pt$-$Cl bond length, ($ii$) a net decrease of the 
N$-$Pt$-$N angles. Analyses of the $g(r)$ associated to the solvent-solute interactions given 
in Figure~\ref{fig:figure1} allows for interpreting the systematic effects in terms of ligand hydration 
shells. 
\begin{table*}[htb!]
\scriptsize{         
	\vspace{10pt}
	\caption{Analysis of the ligand--water $g(r)$ calculated for solvated cisplatin 
          and derivatives. The maxima ($\rmax$) and the integration radii ($\rint$) are given in \AA. 
          The coordination number ($n$) were evaluated by integrating the corresponding $g(r)$ up to $\rint$.
          The atoms labeled $\trm{O}_w$ and $\trm{H}_w$ refer to water molecules. The $\trm{Cl$\cdots$H}_w$ 
          and $\trm{NH$\cdots$O}_w$ RDFs are represented on Figures~\ref{fig:figure1}(a) and (b), respectively.}
	\label{table:2}
	\centerline{
	\begin{tabular}{lccccccccc}
	\hline\hline
Complex &\mcol{2}{c}{\cisPt}  & \mcol{2}{c}{\aquaPt} &\mcol{2}{c}{\aminoPt}&\mcol{2}{c}{\pyriPt}  \\ 
Peak$^a$ & first      & second     & first     & second       & first      & second       & first      & second  \\\hline
    $\trm{Cl$\cdots$H}_w$  &      &            &            &          &               &          &        & \\
	$\rmax$            & 2.3  & (3.9/4.4)* &2.3/2.7\sh  & (4.3/4.9)* & 2.4/2.9\sh  & (5.1)*   &  2.4   &  (4.5/5.1)* \\
	$\rint^b$          & 2.8  &    5.0     &3.0         &  5.0       & 3.1         &  5.0     & 2.9    &   5.0      \\
	$n$	           & 1.8  & 22.7       &2.2         & 22.6       & 2.1         & 19.0     & 1.9    & 19.6   \\\hline
    $\trm{NH$\cdots$O}_w$  &      &       &            &       &      &             &       & \\
	$\rmax$            & 1.9  &  3.4  & 1.9/2.4\sh &  3.5  & 1.9  &  3.5/3.9\sh & 1.9   & 3.4/4.3\sh \\
	$\rint$            & 2.6  &  4.5  & 2.6        &  4.5  & 2.5  &  4.5        & 2.6   & 4.5  \\
	$n$	           & 0.7  &  7.7  & 0.8        &  7.6  & 0.6  &  7.6        & 0.6   & 6.7  \\\hline
\hline   	
\end{tabular}    
}
}
$^a$Asterisks indicate that the peak(s) is(are) not well resolved. Peak shoulder is indicated by \textit{s}.
$^b$For the first integration, $\rint$ was fixed to the closest minimum following the first maximum. 
Beyond these radii fixed values of $\rint=5.0$ and 4.5 \AA\ for the Cl and NH coordination numbers 
respectively, were considered.
\end{table*} 
As already discussed in Ref.~\citen{truflandier_probing_2010} for cisplatin, the NH$\cdots$O$_\trm{w}$ 
$g(r)$ shows that the first and second hydration shells of each NH$_3$ group integrates around 1 and 
8 water molecules, respectively. This also applies to the other platinum complexes (cf. Table~\ref{table:2}).
The fact that the ligand solvation shell is only weakly affected by the variety of ligands bonded to the 
Pt atom and the charge state of the solute is confirmed by the Cl$\cdots$H$_\trm{w}$ $g(r)$ plotted 
on Figure~\ref{fig:figure1}, where each Cl atom integrates in the first neighbor region 
2 waters molecules, independently of the system. The second shell is more difficult to discussed due
to the limited resolution of the peaks. Nevertheless, in light of this results, ($i$) can be directly attributed 
to the Cl$\cdots$H$_{w}$ hydrogen bonds perturbing the Pt$-$Cl bond strength, whereas ($ii$) is an indirect 
consequence of the ammonia group hydration where the preferential arrangement of NH$_3$ from a static 
gas-phase optimisation ---with one of the hydrogen atom pointing towards the closest Cl lone pair--- is lost, 
relaxing the constraint on the Cl$-$Pt$-$N angles of the square planar complexes. Note that, as shown in 
Table~\ref{table:struct_complexes}, a polarizable continuum model (PCM) applied to static isolated solute
is able to simulate these two features, demonstrating the concomitant contributions of specific and non-specific 
solvent effect. In Table~\ref{table:3} is compared selected structural parameters extracted from the RDFs 
for cisplatin and the mono-aqua derivative in regard to the previous studies based on 
DFT-MD.\cite{lau_pccp_2010,truflandier_inorgchem_2011,kroutil_inorgchem_2016} 
Even if the results were obtained from different implementations and exchange-correlation functionals 
---preventing an unbiased comparison--- we observe an overall agreement which, besides validating the 
reliability of our implementation, suggests minor impact of $(i)$ the GGA functional 
(cf. Ref.~\citen{lau_pccp_2010} DFT-MD simulations performed with BLYP) and $(ii)$ inclusion of the empirical 
dispersion correction (cf. Ref.~\citen{kroutil_inorgchem_2016} ; DFT-MD simulations performed with PBE-D3) 
on the qualitative description of the ligand first hydration shells. 

\begin{table*}[ht!]
\scriptsize{         
	\vspace{10pt}
	\caption{Comparison of selected structural parameters (in \AA) calculated from the ligand--water RDFs of solvated cisplatin
	obtained from various DFT-MD simulations. See caption and footnotes of Table~\ref{table:2} for more details.}
	\label{table:3}
	\centerline{
	\begin{tabular}{lcccccccc}
	\hline\hline
	         & \mcol{4}{c}{\cisPt} & & \mcol{3}{c}{\aquaPt} \\ \cmidrule{2-5}\cmidrule{6-9}
Method  & PBE/BOMD
             & PBE/CPMD
             & BLYP/CPMD  
             & PBE+D3/BOMD
             &
             & PBE/BOMD              
             & BLYP/CPMD  
             & PBE+D3/BOMD\\
Basis     & PAO/NCPP
             & PW/USPP
             & PW/NCPP  
             & mixed GTO-PW/NCPP
             &
             & PAO/NCPP              
             & PW/NCPP 
             & mixed GTO-PW/NCPP \\
Reference & this work
             & [\citen{truflandier_inorgchem_2011}]
             & [\citen{lau_hydrolysis_2010}]  
             & [\citen{kroutil_inorgchem_2016}]
             &
             & this work
             & [\citen{lau_hydrolysis_2010}]  
             & [\citen{kroutil_inorgchem_2016}]
              \\\hline 
 $\trm{N$\cdots$O}_w$ &   &    &         &         &$\quad$&         &       &      \\
 $\rmax$            & 3.0    & 3.2*  &  2.9  &  3.0  &$\quad$& 2.9   & 2.9 & 3.0 \\
 $\rint$              & 3.3    & 3.4*  &  --     &  3.6  &$\quad$& 3.7   & --    & 3.5\\ 
 $n$                   & 2.4    & 2.5*  &  2.9  &  3.6  &$\quad$& 3.2   & 3.5 & 3.5\\\hline
 $\trm{NH$\cdots$O}_w$&   &      &            &       &$\quad$&               &    &      \\
 $\rmax$              & 1.9      & 1.9  &   --       &  --   &$\quad$& 1.9/2.4s & -- & -- \\    
 $\rint$                & 2.6      & 2.5   &   --      &  --   &$\quad$& 2.6         & -- & -- \\    
 $n$	                   & 0.7      & 0.7   &   --      &  --   &$\quad$& 0.8         & -- & -- \\\hline
 $\trm{Cl$\cdots$O}_w$ &   &      &            &     &$\quad$&       &       &        \\
 $\rmax$            &   3.4*   & 3.3   &   3.2*   & --  &$\quad$& 3.5 & 3.1 & --  \\
 $\rint$              &   3.7*   & 3.7   &    --      & --  &$\quad$& 3.9 &  --   & --  \\
 $n$                   &   2.8*   & 2.6   &   2.9*   & --  &$\quad$& 3.8 & 3.4 & --  \\\hline
 $\trm{Cl$\cdots$H}_w$ &   &      &            &        &$\quad$&               &       &      \\
 $\rmax$            &  2.3    & 2.3     &   2.3    &  2.3 &$\quad$& 2.3/2.7s & 2.3 & 2.3 \\
 $\rint$              &  2.8     & 2.9     &     --     & 2.8 &$\quad$& 3.0         & --   & 2.9 \\
 $n$                   &  1.8     & 2.2     &   1.7    & 2.2 &$\quad$& 2.2         & 2.4 & 2.2 \\\hline\hline
\end{tabular}    
}
}
\end{table*} 
\begin{figure}[!h]
	\centerline{\includegraphics[width=0.90\textwidth]{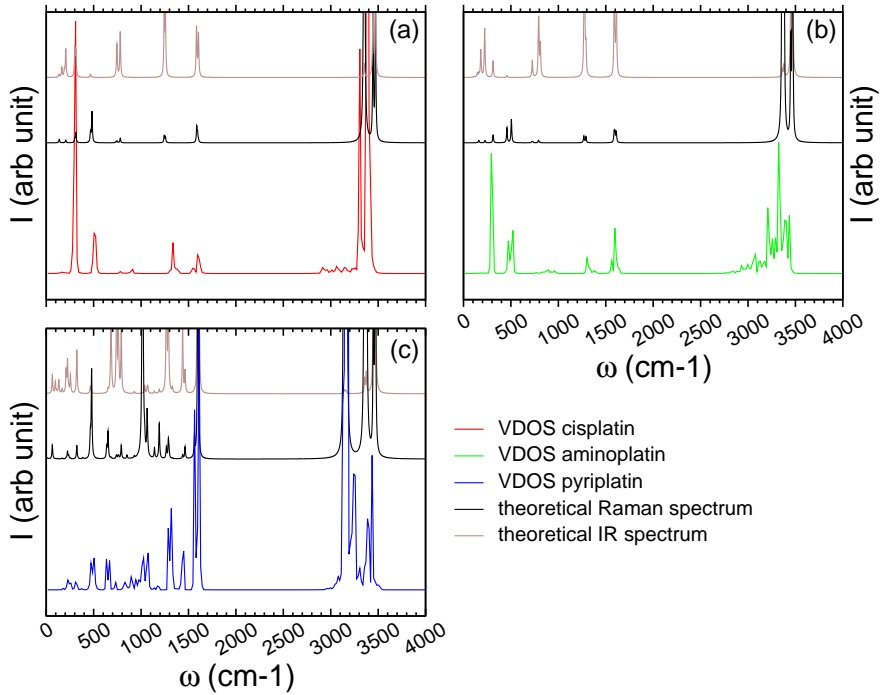}}
	\caption{Comparison between Raman and infra-red theoretical spectra obtained from a
	molecular approach along with an implicit sovent model and the partial vibrational density
	of states (VDOS) extracted from the BOMD simulations.} 
	 \label{fig:figure2}  
\end{figure}
Another important step in the assessment process of our DFT-MD simulations is to ensure that the 
characterisitic vibrational modes of the solute and solvent were properly activated. 
Partial VDOS obtained for cisplatin, aminoplatin and pyriplatin are plotted on Figure~\ref{fig:figure2}.
As external references, Raman and infra-red theoretical spectra of the isolated molecules are also reported.
They were calculated on top of the stationary points obtained at the PBE-GTO-DZP level of theory
including a PCM.\cite{gaussian09_short} Here, it is worth to recall that VDOS heights computed from the Fourier 
transform of the VACFs are not to be compared with spectroscopic intensities, and to emphasize
that even if the underlying semiclassical theory involved in evaluating the VDOS ---based on the
classical propagation of the nuclei on the BO potential energy surface--- is quite different 
from the response function computed from analytic (or numerical) derivatives ---based on the 
harmonic approximation--- their comparisons remain valuable for routine checks. As expected,
cisplatin and aminoplatin vibrational spectrum are similar in many respects since the Pt center is 
surrounded by the same type of ligands. In both cases the bond-stretching modes: N$-$H 
($>$ 3200 cm$^{-1}$), Pt$-$N ($\sim$ 500 cm$^{-1}$) and Pt$-$Cl ($\sim$ 300 cm$^{-1}$) 
are observed as well as the NH$_3$ rocking and symmetric/asymetric 
deformation signatures, found around 800 and 1300/1300 cm$^{-1}$, 
respectively.\cite{nakamoto_infrared_1965,wysokinski_performance_2001,truflandier_inorgchem_2011}
For pyriplatin, the complexity of the vibrational spectra is increased by the contributions
of the pyridine ligand, with the stretching modes: C$-$H ($\sim$ 3250 cm$^{-1}$),
C$-$C/C$-$N (in the range 1100 to 1500 cm$^{-1}$), the skeletal bending modes 
along with the out-of-plane C$-$H wagging ($\sim$ 1000 cm$^{-1}$ and 700 cm$^{-1}$).
As shown on Figure~\ref{fig:figure2}, VDOSs extracted for the DFT-BOMD is in fair accordance
with the theoretical spectra, such that, given the level of theory PAO-DZP-PBE,  
we can be confident in the reliability of our simulations when applied to free energy profiles 
calculation.

\subsection{Inverse hydration}
\label{sec:bomd-analysis}
\begin{figure}[!h]
	\centerline{\includegraphics[width=0.8\textwidth]{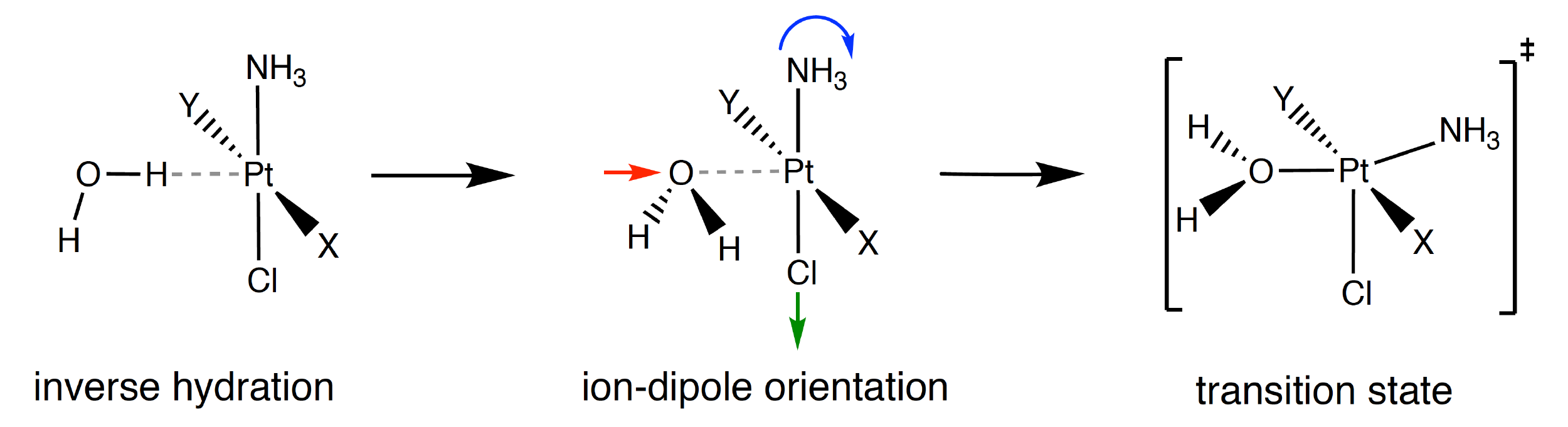}}
	\caption{Structural arrangements of the incoming water molecule at the early stages of the aquation reaction.} 
	 \label{fig:scheme3}  
\end{figure}

\begin{table*}[!h]
\scriptsize{         
	\vspace{10pt}
	\caption{Analysis of the Pt--O and Pt--H RDFs calculated for 
	cisplatin and derivatives along with the coordination numbers $n$ ; cf. caption of
	Table~\ref{table:2} and Figures~\ref{fig:figure1}(c) and (d) for the plots. }
	\label{table:4}	
	\centerline{
	\begin{tabular}{lcccccccc}
	\hline\hline
  Complex      & \mcol{2}{c}{\cisPt}   & \mcol{2}{c}{\aquaPt}    & \mcol{2}{c}{\aminoPt}    & \mcol{2}{c}{\pyriPt}      \\
   Peak          & first        & second  & first        & second   &  first    & second       & first       & second \\\hline
$\trm{Pt$\cdots$H}_w$ &    &         &     &           &     &        &      &           \\ 
	$\rmax$   & 2.5   & 3.0     & 2.4 & (2.8/3.2) & 2.4 &    na  & 2.3  &  (3.0/4.0) \\
	$\rint$     & 2.6   & 3.3     & 2.6 & 3.3       & 2.6 &   3.3  & 2.6  &  3.3       \\
	$n^a$      & 0.5   & 1.4     & 0.2 & 0.9       & 0.2 &   0.7  & 0.1  &  0.7       \\\hline
$\trm{Pt$\cdots$O}_w$  &   &      &     &      &      &           &      &  \\
	$\rmax$   & 3.5   & 4.3  & na  & 4.2  & 3.4  & (4.2/4.5) & 3.3  & (3.9/4.6)  \\        
 	$\rint$     & 3.6   & 4.6  & 3.6 & 4.6  & 3.6  &  4.6      & 3.6  &  4.6       \\
	$n^a$      & 0.7   & 6.4  & 0.7 & 7.0  & 0.5  &  6.7      & 0.4  &  5.3       \\
\hline\hline
\end{tabular}
}
}
$^a$Since RDFs are poorly-resolved, fixed $\rint$ values were considered for the first and second integration.
\end{table*}
We shall now focussed on the inverse-hydration taking place in the axial region of the
square-planar Pt(II) complexes (Scheme~\ref{fig:scheme3}). It can be observed
at the beginning of each MD movies provided in the SI. The hydrogen-like 
bonding between a water molecule and the metal center following a H-ahead orientation 
is easily revealed by analysing the \gPtO\ and \gPtH\ available on Figure~\ref{fig:figure1}. 
First we note that for all the systems investigated in this work a non-negligible probability 
of having a Pt$\cdots$(H$_2$O)$_\trm{w}$ contact is observed at an intermolecular distance 
of around 2.4 \AA. The poorly-resolved peaks obtained for the Pt$\cdots$H$_\trm{w}$ 
(Pt$\cdots$O$_\trm{w}$) $g(r)$ in the range 2 to 3 (3 to 4) \AA\ illustrate the motion 
or/and the exchange of H$_2$O with other solvent molecules in close proximity, eg. those 
available from the Cl hydration shell(s). Another important insight is brought by the height of the 
peaks, cf. the Pt$\cdots$H$_\trm{w}$ (Pt$\cdots$O$_\trm{w}$) $g(r)$ around 2.4 (3.4)
 \AA, indicating a net decrease of the contact occurrence when going from neutral to positively
charged complexes. Quantification of the Pt$\cdots$H$_2$O contact is provided in Table~\ref{table:4}
for the bi- and monofunctional platinum anticancer drugs, through the evaluation 
of the Pt(H) and Pt(O) coordination numbers, noted \nPtH\ and \nPtO, respectively.
Due to the broad shape of the peaks, a set of two integration radius ($\rint$ in Table~\ref{table:3}) 
were used independently of the complex.
For neutral cisplatin, \nPtH\ is evaluated to be within the interval $[0.5,1.4]$,
along with $[0.7,6.4]$ for \nPtO. Note that for larger $\rint$, it is likely that the value of the coordination numbers 
also incorporate H$_2$O molecules from the NH$_3$ and Cl ligand solvent shells. For the set of positively charged complexes, 
the probability of Pt$\cdots$(H$_2$O)$_\trm{w}$ contact is significantly reduced from $n_\trm{PtH}=[0.2,0.9]$ 
for mono-aqua cisplatin to $n_\trm{PtH}=[0.1,0.7]$ for pyriplatin. For this set, variations of $n_\trm{PtO}$ 
seems to be more affected by the type of ligand, for which, mono-aqua cisplatin $n_\trm{PtO}$ 
presents values similar to its neutral parent, whereas for aminoplatin and pyriplatin the coordination
number is lowered to $[0.5,6.7]$ and $[0.4,5.3]$, respectively. These results corroborate the work of Kroutil 
et al.\cite{kroutil_inorgchem_2016} which showed, among other findings, that: (i) H$_\trm{w}$ 
coordination number decreases with increasing positive charge of the complex, 
(ii) given a state charge (+1 in our case) the H-ahead orientation toward the Pt atom is 
strongly dependent on the hydrogen bond network formed by the nearby ligands.
We refer the reader to this reference for extended discussion on this topic which
also includes energetics consideration.

Concerning now the second part of this work dealing with evaluation of the hydrolysis free 
energy profiles of reaction (\textbf{1}) and (\textbf{2}), among all the water molecules 
accessible within the solute first solvation shell, it seems quite reasonable to consider the 
inversely bounded water molecule as the reactant.

\subsection{Hydrolysis free energy profiles}
The choice of the reaction coordinate $\xi$ is an important point in BME
integration which can be a delicate issue when little is known about the chemical
reaction. In this work, we have relied on the broad (experimental and theoretical)
litterature agreeing that cisplatin aquation(s) is a one step process, eg. the S$_{N}$2-like 
mechanism, for which the reactants (R) and products (P) are connected by a single 
transition state (TS). From this assumption, a rapid analysis of the corresponding saddle 
point (optimized with a quantum chemistry code) shows that the characteristic imaginary 
frequency is related to the Cl$-$Pt$-$O antisymmetric stretching mode together
with the in-plane rotation of the NH$_3$ group in trans position (Scheme~\ref{fig:scheme3}). 
This led us to consider the difference of distance\cite{raugei_ab_1999,yang2004free,buhl_mechanism_2006,komeiji_fragment_2009} 
for the reaction coordinate:
\begin{equation}
	\xi = ||{\bf r}_\trm{Pt} - {\bf r}_{\trm{O}}|| -  ||{\bf r}_\trm{Pt} - {\bf r}_\trm{Cl}||,
\end{equation}
where O is the oxygen atom of the H$\cdots$Pt bonded water molecule. Configurations used 
to initiate the constrained DFT-BOMD trajectories were extracted from the canonical ensemble simulations 
discussed in Sec.~\ref{sec:bomd-analysis}. Note that variations of (Helmholtz) free energies as computed 
from Eq.~(\ref{eq:thermo}) were directly converted to (Gibbs) free enthalpies since we did not observe 
any strong variation of pressure during the thermodynamic integration which would have
significantly modified the final enthalpy values.
\begin{figure}[!htb]
	\centerline{\includegraphics[width=0.8\textwidth]{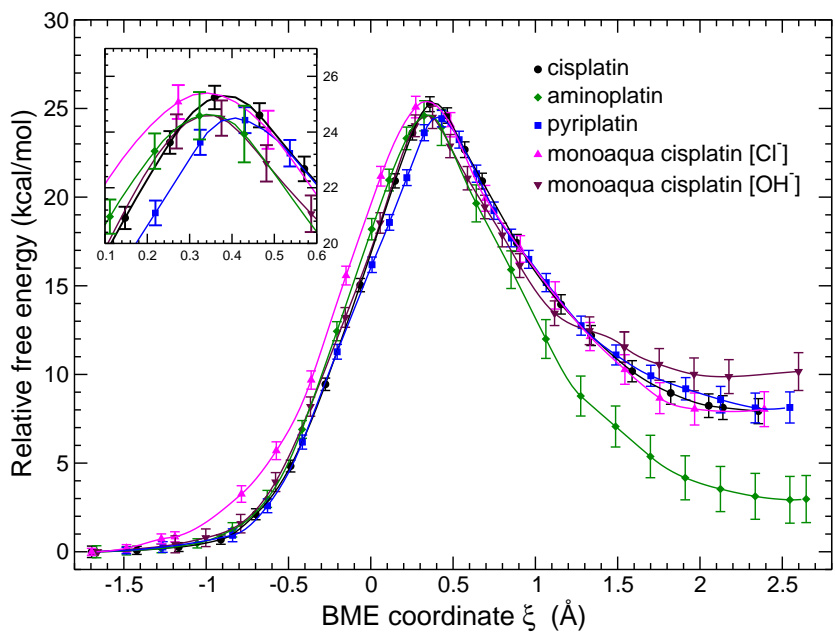}}
	\caption{Free energy path for hydrolysis of cisplatin and derivatives in water at $T = 310$ K
	using the blue-moon ensemble approach. Curves were interpolated with cubic splines.} 
	 \label{fig:figure4}  
\end{figure}

Free energy reaction paths are displayed in Figure~\ref{fig:figure4} along with the free enthalpies of activation 
$(\Delta^\ddagger G)$ and reaction $(\Delta_rG)$ collected in Table~\ref{tab:table1}. Concerning the second 
hydrolysis of cisplatin performed \textit{via} the monoaqua complex, two set of BME calculation were 
produced: the first with Cl$^-$ as counterion following the first hydrolysis (later abbreviated by monoaqua[Cl]), 
the second from afresh \textit{NVT} equilibration in which Cl$^-$ has been replaced by OH$^-$ 
(abbreviated by monoaqua[OH]).  From Figure~\ref{fig:figure4} we observe very similar 
energy profiles from the attack of the water molecule up to the transitions states. Evaluated $\Delta^\ddagger G$
for cationic monofuntional species are found to be lower of about 1 ${\rm kcal}\cdot {\rm mol}^{-1}$ 
than the reference value of 25 ${\rm kcal}\cdot {\rm mol}^{-1}$ obtained for the neutral parent. The 
presence of the Cl$^-$ counterion during the second hydrolysis of cisplatin tends to slightly decrease the
barrier, but without dramatic effect. Indeed, major deviations between aquation profiles are found after the TS.
This lead to noticeable differences in the values of $\Delta_rG$,
for which we found that the aminoplatin aquation is endothermic by 3 ${\rm kcal}\cdot {\rm mol}^{-1}$ 
compared to 8 for cisplatin, pyriplatin and moaqua[Cl], whereas for moaqua[OH] the free enthalpy of reaction reach
the value of 10 ${\rm kcal}\cdot {\rm mol}^{-1}$. It is tempting to compare these values to available 
experimental data. This should be done with lots of care owing to: $(i)$ the large amount of papers dealing with
this subject, which comes generally with the same amount of differences in experimental condition, $(ii)$
the limitations of our model in reproducing these conditions. 
\begin{table}[!htb]
\begin{tabular}{l|lcc}
\hline
\hline
Reactant & Counterion & $\Delta^\ddagger G$ $({\rm kcal}\cdot {\rm mol}^{-1})$ &  
$\Delta_rG$ $({\rm kcal}\cdot {\rm mol}^{-1}$)\\
\hline
cisplatin                  & none        & $25.3 \pm 0.4$ &  $7.9   \pm 0.7$ \\
mono-aqua cisplatin & OH$^-$    & $25.4 \pm 0.6$ & $10.2 \pm 1.1$ \\
                              & Cl$^-$      & $24.5 \pm 0.6$ & $8.0 \pm 1.0$ \\
aminoplatin             & OH$^-$     & $24.5 \pm 0.8$ & $2.9 \pm 1.3$\\
pyriplatin                & OH$^-$     & $24.4 \pm 0.5$ & $8.1 \pm 0.9$ \\
\hline
\hline
\end{tabular}
\caption{Free enthalpies of activations and reactions for aquation of cisplatin and derivatives in water at $T = 310$ K
using the blue-moon ensemble approach.}
\label{tab:table1} 
\end{table}
As a result, even if the pH of the aqueous media can impact significantly the kinetic of the reactions 
---as shown in the comprehensive study of House and coworkers\cite{miller_hydrolysis_1989_1,miller_hydrolysis_1989_2,miller_hydrolysis_1990}--- 
we must emphasize that the precision reach by the BME simulations remains far from the chemical accuracy 
required to discuss variation in activation energies below 1 ${\rm kcal}\cdot {\rm mol}^{-1}$.\footnote{It is
worth to recall that a variation of 1 ${\rm kcal}\cdot {\rm mol}^{-1}$ on the free energy of activation convert 
roughly to a variation of one order of magnitude on the rate constant.} From Refs.~\citen{miller_hydrolysis_1989_1,miller_hydrolysis_1989_2,reishus_cisplatin_1961,marti_reversible_1998,davies_slowing_2000,lee_cisplatin_1976,vinje_influence_2005} 
we can establish a consensus on the experimental free enthalpy (at ambient temperature) for both, the first
and second aquation, to be between 23 and 24 ${\rm kcal}\cdot {\rm mol}^{-1}$. The experimental values
for the cationic aminoplatin and pyriplatin are expected to be within this range.\cite{park_pnas_2012}
A consensus on the endothermic property of the aquation reactions can also be drawn, with
$\Delta_rG$ for cisplatin experimentally evaluated to be in the range 3-10 ${\rm kcal}\cdot {\rm mol}^{-1}$.\cite{miller_hydrolysis_1989_1,miller_hydrolysis_1989_2,lee_cisplatin_1976} Our results which are in line with 
experimental data should not be overinterpreted. Whereas we are quite confident in the reliability of the computed
BME activation enthalpies, the thermodynamic quantities must not be considered as definitive. The limited size 
of cell can not prevent intramolecular ion pair contact to occur (the outermost Cl$\cdots$Pt distance being of 6.5 \AA).
For aminoplatin, the larger error bars on the integrated free energy observed at the late stages of the reaction
suggest that the final $\Delta_rG$ is not fully converged.

\begin{figure}[!htb]
	\centerline{\includegraphics[width=1.0\textwidth]{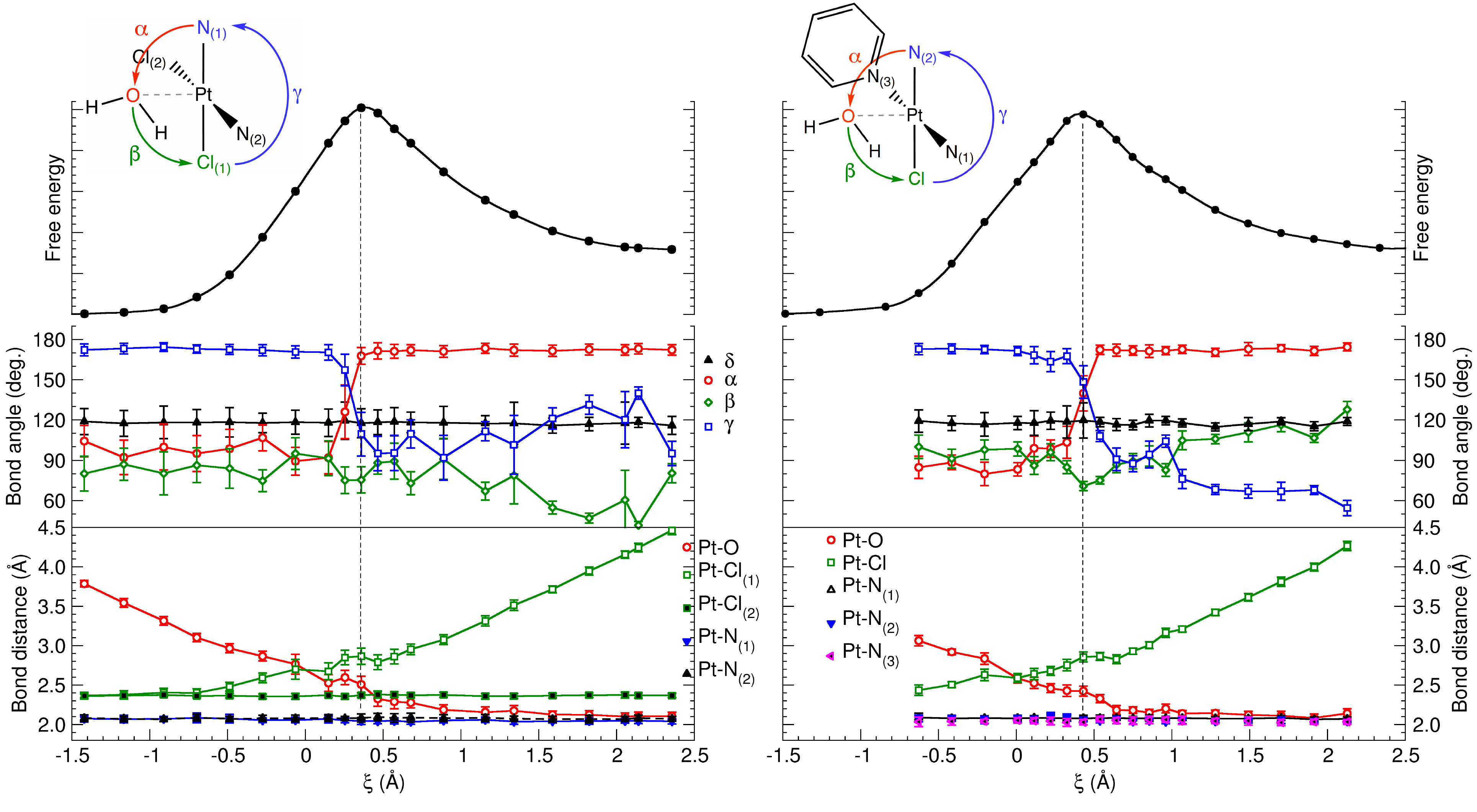}}
	\caption{Ensemble average of the skeletal bond angles and distances as a function of the 
	BME coordinate obtained for cisplatin (left) and pyriplatin (right) aquation. Vertical bars
	correspond to the standard deviations. Free energy profile along the reaction path has been 
	included to guide the eye.}
	 \label{fig:figure5}  
\end{figure}

\begin{figure}[!htb]
	\centerline{\includegraphics[width=1.0\textwidth]{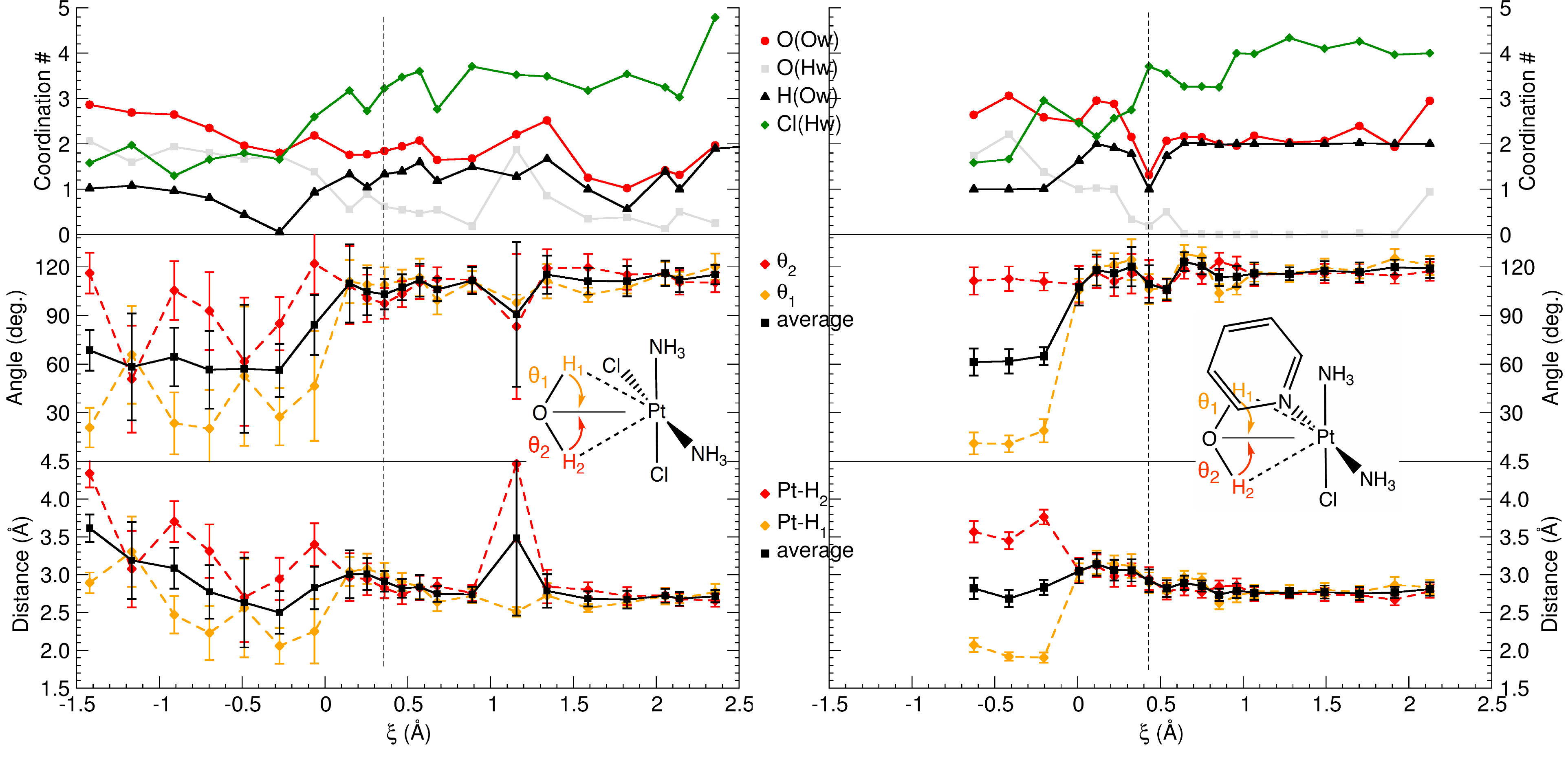}}
	\caption{Ensemble average of selected geometrical parameters describing the attack of 
	inversely bonded water molecule for cisplatin (left) and pyriplatin (right) aquation reaction. 
	See also the caption of Figure~\ref{fig:figure5}.  Coordination numbers reported on the upper 
	panel were evaluated from the RDFs.}
	 \label{fig:figure6}  
\end{figure}

In order to gain some insights into the reactants motion and solvent reorganisation along the reaction 
path, visual inspection of the constrained MD movies is quite instructive. Representative examples 
are provided in the SI for cisplatin and pyriplatin. For all the complexes investigated similar trends in
the evolution of skeletal bond parameters and solvent shell relative to the reactant constrained dynamics were 
found. The study cases of cisplatin and pyriplatin aquations are depicted in Figures~\ref{fig:figure5} 
and~\ref{fig:figure6}, where the variation of selected structural parameters are plotted as a function 
of $\xi$. At the beginning of the reaction ($\xi < 0$), the square planar structure with 
the inverse-hydration is globally conserved in both cases. In this regime, for cisplatin, the net
decrease of the Pt$\cdots$O bond length ($d_\trm{PtO}=3.8\rightarrow2.7$ \AA) is accompanied 
by a weaker increase of the Pt$\cdots$Cl distance ($d_\trm{PtCl}=2.4\rightarrow2.7$ \AA) 
whereas the N$-$Pt$-$O ($\alpha$), O$-$Pt$-$Cl ($\beta$) and Cl$-$Pt$-$N ($\gamma$ ) angles 
oscillate around mean values of 100, 85 and 175$^\circ$, respectively. These observations also
hold for pyriplatin with slight differences in values. For cisplatin, magnitude of the variations
along $\xi$ observed for $\alpha$ and $\beta$, which are much higher than for $\gamma$, indicates 
that the constrained water molecule remains free to visit the axial region of the square-planar complex.
It seems to be less pronounced for pyriplatin. The difference in lability of the inversely 
bonded H$_2$O between the two complexes is apparent when comparing the standard deviation
(std) of the Pt$\cdots$H distances plotted on Figure~\ref{fig:figure6}. For cisplatin we occasionally 
observe that these distances collapse towards a single value, eg. at $\xi=(-1.2,-0.5)$, 
indicating H-atom swap through the rotation along the axis perpendicular to the H$_2$O 
molecular plan. The variations of the H$-$O$-$Pt angles $(\theta_1,\theta_2)$ with mean 
values of about $(20,100)\pm 20^\circ$ for cisplatin, and $(10,110)\pm 5^\circ$ for pyriplatin 
confirm the preferential inverse-hydration orientation, and for the later case, the reduced mobility
of the H$_2$O-reactant. 

When approaching the transition state located at $\xi=0.36$ and $0.43$ \AA\ for cisplatin 
and pyriplatin respectively, the inversely bonded H$_2$O is activated \textit{via} the ion-dipole 
orientation depicted in Scheme~\ref{fig:scheme3} with $d_\trm{PtO}\simeq 2.5$ and $d_\trm{PtH}\simeq 3.0$ 
\AA. For cisplatin, when $\xi>0.1$ ($\xi>0.3$ for pyriplatin), the rotation of the NH$_3$ group is 
engaged leading to the TS. Note that, within the TS region, reported Pt$\cdots$H distances 
and angles are weakly affected compared to the rotation angles 
$(\alpha,\gamma)$. By identifying on Figure~\ref{fig:figure5} the skeletal parameters corresponding 
to the maximum of free energy, it seems that the cisplatin TS is closer to a square pyramidal
structure than a trigonal bipyramid as observed in gas phase 
calculations.\cite{zhang_hydrolysis_2001,
*robertazzi_hydrogen_2004,         *lau_hydrolysis_2006,
*alberto_second-generation_2009,*lucas_neutral_2009,
*banerjee_cpl_2010,*melchior_comparative_2010} 
With $\alpha\simeq\gamma\simeq 140^\circ$ a trigonal pyramid can be clearly assigned
to pyriplatin. For $\xi>0.9$, the water molecule is fully coordinated to the metal center with a typical 
bond length $d_\trm{PtO}\simeq 2.1$ \AA, and the Cl anion is released to bulk water. Inspection of 
the cisplatin BME integration movie at $\xi=1.15$ reveals that the singularity on Figure~\ref{fig:figure6} 
is due to the departure of one of the H$_2$O-ligand proton towards the solvent shell. 
From Figure~\ref{fig:figure5}, looking to the Pt$-$N and unconstrained Pt$-$Cl mean bond lengths, 
it is interesting to outline that the corresponding ligands are spectator of the reaction
in agreement with the in-plane symmetry of the transition state. This is well exemplified by the
mean angle $\delta=(\alpha+\beta+\gamma)/3$ which remains fixed to 120$^\circ$ 
with standard deviation not exceeding 10$^\circ$ for both cases, showing that there are only weak 
perturbations from outside the N$-$O$-$Cl plane.

By visualizing the constrained MD trajectories reported in the SI, it is rather clear  that the activated 
water molecule is embedded in a hydrogen-bond network throughout the whole course of the 
aquation reaction. To quantify the number of first nearest neighbor solvent molecules surrounding 
the H$_2$O-reactant and Cl-product, the coordination numbers \nOOw, (\nOHw,\nHOw) and 
\nClHw\ are plotted on Figure~\ref{fig:figure6} as a function of $\xi$. They were evaluated from 
the ensemble averages (cf. Sec.~\ref{sec:bme-method}) by integrating the corresponding RDFs 
up to standard radii\cite{zhang_chlorine_2013,distasio_JCP_2014} of 3.2, 2.5 and 2.9 \AA, 
respectively. Even if at first sight the evolution of $n$ is somewhat erratic due to the short time 
sampling, we can extract some relevant trends. At the beginning of the reaction, the inversely
bonded water molecule integrates around 3 solvent molecules respecting its H-bond 
donor[1]/acceptor[2] capacity. Around the ion-dipole orientation region, as expected, 
H-bond acceptor (donor) capacity of the H$_2$O-reactant reduces (raises). At the 
transition state, around 2 and 1 solvent molecule(s) are in close proximity of the activated 
H$_2$O in cisplatin and pyriplatin, respectively. After this step, the average number
acceptor-type H-bond continue to diminish whereas donor-type stabilizes to 2. The evolution
of the H$_2$O-reactant H-bond network contrasts with the global increase of the solvent shell
size surrounding the Cl-product. Due the limited size of the model \nClHw\ never 
really stabilized to the reference value comprised between 5 and 6 as obtained for free
solvated chlorine anion.\cite{zhang_chlorine_2013} 

Finally, the BME aquation profiles and their analysis suggest that the aquation reaction 
of cisplatin and its monofunctional derivatives presenting a Cl-leaving group is likely to be independent of the 
charge state the type and number of N-donor ligands despite variations in the mobility and microsolvation of the 
activated water molecule. Even if conditioned by the choice of coordinate, the inverse hydration followed by the 
ion-dipole orientation seems to be a prerequisite to reach the transition state. 

\section{Conclusion and Outlook}
\label{sec:conclusion}

In this work, reliable DFT-MD simulations of the aquation process of cisplatin and a set of monofunctional
platinum anticancer drugs have been performed using constrained dynamics in conjuction with BME 
thermodynamic integration. Given the level of theory, accuracy of about 1 ${\rm kcal}\cdot {\rm mol}^{-1}$
has been reached on the activation barrier. For the set of square-planar Pt(II) complexes, it is found that the kinetic 
of the reaction is independent of the ligands and the charge state. It seems that the free enthalpies of reaction
differ from a few ${\rm kcal}\cdot {\rm mol}^{-1}$ from one complex to another, and counterions can impact 
the thermodynamic quantity with variations in the same range of magnitude. This result need to be confirmed
by performing larger scale DFT-MD simulations. We conclude that it is likely that the mono and bifunctional
platinum anticancer drugs behave the same way during the time lapse between the cellular uptake
and the DNA platination.

From a computational point of view, such kind of approach alleviates many intricacies related to modelling
chemical reactions in aqueous solvent which are commonly based on isolated molecule approximation 
where thermodynamic quantities are computed from the ideal gas approximation. Even if the protocol
used here is computationally expensive, we believe that it will become a routine task in near future along
with the possibility of increasing the system size and using more sophisticated exchange correlation functionals,
especially with the \textsc{Conquest} code which have been developped to perform large scale DFT simulations.
This work opens the way for futur high level theoretical investigations of the anticancer activity of monofuntional 
platinum complexes which are important steps for approaching \textit{in silico} design of new 
potent drugs.

\providecommand{\latin}[1]{#1}
\providecommand*\mcitethebibliography{\thebibliography}
\csname @ifundefined\endcsname{endmcitethebibliography}
  {\let\endmcitethebibliography\endthebibliography}{}




\begin{mcitethebibliography}{90}
\providecommand*\natexlab[1]{#1}
\providecommand*\mciteSetBstSublistMode[1]{}
\providecommand*\mciteSetBstMaxWidthForm[2]{}
\providecommand*\mciteBstWouldAddEndPuncttrue
  {\def\EndOfBibitem{\unskip.}}
\providecommand*\mciteBstWouldAddEndPunctfalse
  {\let\EndOfBibitem\relax}
\providecommand*\mciteSetBstMidEndSepPunct[3]{}
\providecommand*\mciteSetBstSublistLabelBeginEnd[3]{}
\providecommand*\EndOfBibitem{}
\mciteSetBstSublistMode{f}
\mciteSetBstMaxWidthForm{subitem}{(\alph{mcitesubitemcount})}
\mciteSetBstSublistLabelBeginEnd
  {\mcitemaxwidthsubitemform\space}
  {\relax}
  {\relax}

\bibitem[Wheate \latin{et~al.}(2010)Wheate, Walker, Craig, and
  Oun]{wheate_status_2010}
Wheate,~N.~J.; Walker,~S.; Craig,~G.~E.; Oun,~R. The status of platinum
  anticancer drugs in the clinic and in clinical trials. \emph{Dalton Trans.}
  \textbf{2010}, \emph{39}, 8113--8127\relax
\mciteBstWouldAddEndPuncttrue
\mciteSetBstMidEndSepPunct{\mcitedefaultmidpunct}
{\mcitedefaultendpunct}{\mcitedefaultseppunct}\relax
\EndOfBibitem
\bibitem[Johnstone \latin{et~al.}(2014)Johnstone, Park, and
  Lippard]{johnstone_anticancer_2014}
Johnstone,~T.~C.; Park,~G.~Y.; Lippard,~S.~J. Understanding and {Improving}
  {Platinum} {Anticancer} {Drugs} {\textendash} {Phenanthriplatin}.
  \emph{Anticancer Res.} \textbf{2014}, \emph{34}, 471--476\relax
\mciteBstWouldAddEndPuncttrue
\mciteSetBstMidEndSepPunct{\mcitedefaultmidpunct}
{\mcitedefaultendpunct}{\mcitedefaultseppunct}\relax
\EndOfBibitem
\bibitem[Apps \latin{et~al.}(2015)Apps, Choi, and Wheate]{apps_Ptdrugs_2015}
Apps,~M.~G.; Choi,~E. H.~Y.; Wheate,~N.~J. The state-of-play and future of
  platinum drugs. \emph{Endocr. Relat. Cancer} \textbf{2015}, \emph{22},
  R219--R233\relax
\mciteBstWouldAddEndPuncttrue
\mciteSetBstMidEndSepPunct{\mcitedefaultmidpunct}
{\mcitedefaultendpunct}{\mcitedefaultseppunct}\relax
\EndOfBibitem
\bibitem[Johnstone \latin{et~al.}(2016)Johnstone, Suntharalingam, and
  Lippard]{johnstone_chemrev_2016}
Johnstone,~T.~C.; Suntharalingam,~K.; Lippard,~S.~J. The {Next} {Generation} of
  {Platinum} {Drugs}: {Targeted} {Pt}({II}) {Agents}, {Nanoparticle}
  {Delivery}, and {Pt}({IV}) {Prodrugs}. \emph{Chem. Rev.} \textbf{2016},
  \emph{116}, 3436--3486\relax
\mciteBstWouldAddEndPuncttrue
\mciteSetBstMidEndSepPunct{\mcitedefaultmidpunct}
{\mcitedefaultendpunct}{\mcitedefaultseppunct}\relax
\EndOfBibitem
\bibitem[Brabec and Kasparkova(2005)Brabec, and
  Kasparkova]{brabec_resistance_2005}
Brabec,~V.; Kasparkova,~J. Modifications of {DNA} by platinum complexes.
  \emph{Drug Resistance Updates} \textbf{2005}, \emph{8}, 131--146\relax
\mciteBstWouldAddEndPuncttrue
\mciteSetBstMidEndSepPunct{\mcitedefaultmidpunct}
{\mcitedefaultendpunct}{\mcitedefaultseppunct}\relax
\EndOfBibitem
\bibitem[Chabner and Roberts(2005)Chabner, and Roberts]{chabner_nature_2005}
Chabner,~B.~A.; Roberts,~T.~G. Chemotherapy and the war on cancer. \emph{Nat.
  Rev. Cancer} \textbf{2005}, \emph{5}, 65--72\relax
\mciteBstWouldAddEndPuncttrue
\mciteSetBstMidEndSepPunct{\mcitedefaultmidpunct}
{\mcitedefaultendpunct}{\mcitedefaultseppunct}\relax
\EndOfBibitem
\bibitem[Kelland(2007)]{kelland_nature_2007}
Kelland,~L. The resurgence of platinum-based cancer chemotherapy. \emph{Nat.
  Rev. Cancer} \textbf{2007}, \emph{7}, 573--584\relax
\mciteBstWouldAddEndPuncttrue
\mciteSetBstMidEndSepPunct{\mcitedefaultmidpunct}
{\mcitedefaultendpunct}{\mcitedefaultseppunct}\relax
\EndOfBibitem
\bibitem[Hambley(1997)]{hambley_coordchemrev_1997}
Hambley,~T.~W. The influence of structure on the activity and toxicity of {Pt}
  anti-cancer drugs. \emph{Coord. Chem. Rev.} \textbf{1997}, \emph{166},
  181--223\relax
\mciteBstWouldAddEndPuncttrue
\mciteSetBstMidEndSepPunct{\mcitedefaultmidpunct}
{\mcitedefaultendpunct}{\mcitedefaultseppunct}\relax
\EndOfBibitem
\bibitem[Ziegler \latin{et~al.}(2000)Ziegler, Silverman, and
  Lippard]{ziegler_jbic_2000}
Ziegler,~C.~J.; Silverman,~A.~P.; Lippard,~S.~J. High-throughput synthesis and
  screening of platinum drug candidates. \emph{J. Biol. Inorg. Chem.}
  \textbf{2000}, \emph{5}, 774--783\relax
\mciteBstWouldAddEndPuncttrue
\mciteSetBstMidEndSepPunct{\mcitedefaultmidpunct}
{\mcitedefaultendpunct}{\mcitedefaultseppunct}\relax
\EndOfBibitem
\bibitem[Jamieson and Lippard(1999)Jamieson, and
  Lippard]{jamieson_chemrev_1999}
Jamieson,~E.~R.; Lippard,~S.~J. Structure, {Recognition}, and {Processing} of
  {Cisplatin}-{DNA} {Adducts}. \emph{Chem. Rev.} \textbf{1999}, \emph{99},
  2467--2498\relax
\mciteBstWouldAddEndPuncttrue
\mciteSetBstMidEndSepPunct{\mcitedefaultmidpunct}
{\mcitedefaultendpunct}{\mcitedefaultseppunct}\relax
\EndOfBibitem
\bibitem[Dhar and Lippard(2009)Dhar, and Lippard]{dhar_review_2009}
Dhar,~S.; Lippard,~S.~J. In \emph{Platinum and {Other} {Heavy} {Metal}
  {Compounds} in {Cancer} {Chemotherapy}}; Bonetti,~A., Leone,~R.,
  Muggia,~F.~M., Howell,~S.~B., Eds.; Cancer {Drug} {Discovery} and
  {Development}; Humana Press, 2009; pp 135--147\relax
\mciteBstWouldAddEndPuncttrue
\mciteSetBstMidEndSepPunct{\mcitedefaultmidpunct}
{\mcitedefaultendpunct}{\mcitedefaultseppunct}\relax
\EndOfBibitem
\bibitem[Vidossich \latin{et~al.}(2016)Vidossich, Lled{\'o}s, and
  Ujaque]{vidossich_accchemressaccounts_2016}
Vidossich,~P.; Lled{\'o}s,~A.; Ujaque,~G. First-{Principles} {Molecular}
  {Dynamics} {Studies} of {Organometallic} {Complexes} and {Homogeneous}
  {Catalytic} {Processes}. \emph{Acc. Chem. Res.} \textbf{2016}, \emph{49},
  1271--1278\relax
\mciteBstWouldAddEndPuncttrue
\mciteSetBstMidEndSepPunct{\mcitedefaultmidpunct}
{\mcitedefaultendpunct}{\mcitedefaultseppunct}\relax
\EndOfBibitem
\bibitem[Beret \latin{et~al.}(2008)Beret, Mart{\'i}nez, Pappalardo,
  {S{\'a}nchez Marcos}, Doltsinis, and Marx]{beret_jctc_2008}
Beret,~E.~C.; Mart{\'i}nez,~J.~M.; Pappalardo,~R.~R.; {S{\'a}nchez Marcos},~E.;
  Doltsinis,~N.~L.; Marx,~D. Explaining Asymmetric Solvation of {Pt(II)} versus
  {Pd(II)} in Aqueous Solution Revealed by Ab Initio Molecular Dynamics
  Simulations. \emph{J. Chem. Theory Comput.} \textbf{2008}, \emph{4},
  2108--2121\relax
\mciteBstWouldAddEndPuncttrue
\mciteSetBstMidEndSepPunct{\mcitedefaultmidpunct}
{\mcitedefaultendpunct}{\mcitedefaultseppunct}\relax
\EndOfBibitem
\bibitem[Beret \latin{et~al.}(2009)Beret, Pappalardo, Marx, and
  S{\'a}nchez~Marcos]{beret_cpc_2009}
Beret,~E.~C.; Pappalardo,~R.~R.; Marx,~D.; S{\'a}nchez~Marcos,~E.
  Characterizing {Pt}-{Derived} {Anticancer} {Drugs} from {First} {Principles}:
  {The} {Case} of {Oxaliplatin} in {Aqueous} {Solution}. \emph{ChemPhysChem}
  \textbf{2009}, \emph{10}, 1044--1052\relax
\mciteBstWouldAddEndPuncttrue
\mciteSetBstMidEndSepPunct{\mcitedefaultmidpunct}
{\mcitedefaultendpunct}{\mcitedefaultseppunct}\relax
\EndOfBibitem
\bibitem[Vidossich \latin{et~al.}(2011)Vidossich, Ortu{\~n}o, Ujaque, and
  Lled{\'o}s]{vidossich_chemphyschem_2011}
Vidossich,~P.; Ortu{\~n}o,~M.~{\'A}.; Ujaque,~G.; Lled{\'o}s,~A. Do
  {Metal}...{Water} {Hydrogen} {Bonds} {Hold} in {Solution}? {Insight} from
  {Ab} {Initio} {Molecular} {Dynamics} {Simulations}. \emph{ChemPhysChem}
  \textbf{2011}, \emph{12}, 1666--1668\relax
\mciteBstWouldAddEndPuncttrue
\mciteSetBstMidEndSepPunct{\mcitedefaultmidpunct}
{\mcitedefaultendpunct}{\mcitedefaultseppunct}\relax
\EndOfBibitem
\bibitem[Truflandier \latin{et~al.}(2011)Truflandier, Sutter, and
  Autschbach]{truflandier_inorgchem_2011}
Truflandier,~L.~A.; Sutter,~K.; Autschbach,~J. Solvent {Effects} and {Dynamic}
  {Averaging} of $^{195}$Pt {NMR} {Shielding} in {Cisplatin} {Derivatives}.
  \emph{Inorg. Chem.} \textbf{2011}, \emph{50}, 1723--1732\relax
\mciteBstWouldAddEndPuncttrue
\mciteSetBstMidEndSepPunct{\mcitedefaultmidpunct}
{\mcitedefaultendpunct}{\mcitedefaultseppunct}\relax
\EndOfBibitem
\bibitem[Sutter \latin{et~al.}(2011)Sutter, Truflandier, and
  Autschbach]{sutter_chemphyschem_2011}
Sutter,~K.; Truflandier,~L.~A.; Autschbach,~J. {NMR} {J}-{Coupling} {Constants}
  in {Cisplatin} {Derivatives} {Studied} by {Molecular} {Dynamics} and
  {Relativistic} {DFT}. \emph{ChemPhysChem} \textbf{2011}, \emph{12},
  1448--1455\relax
\mciteBstWouldAddEndPuncttrue
\mciteSetBstMidEndSepPunct{\mcitedefaultmidpunct}
{\mcitedefaultendpunct}{\mcitedefaultseppunct}\relax
\EndOfBibitem
\bibitem[Baidina \latin{et~al.}(1981)Baidina, Podberezskaya, Krylova, and
  Borisov]{baidina_crystal_1981}
Baidina,~I.~A.; Podberezskaya,~N.~V.; Krylova,~L.~F.; Borisov,~S.~V. Crystal
  structure of trans-dichloroammineglycineplatinum ({II}) monohydrate
  trans-[{PtNH}$_3$({H}$_2$NCH$_2$COOH){Cl}$_2$]{\textperiodcentered}{H}$_2$O. \emph{J.
  Struct. Chem.} \textbf{1981}, \emph{22}, 463--465\relax
\mciteBstWouldAddEndPuncttrue
\mciteSetBstMidEndSepPunct{\mcitedefaultmidpunct}
{\mcitedefaultendpunct}{\mcitedefaultseppunct}\relax
\EndOfBibitem
\bibitem[Rizzato \latin{et~al.}(2010)Rizzato, Berg{\`es}, Mason, Albinati, and
  Kozelka]{rizzato_angew_2010}
Rizzato,~S.; Berg{\`es},~J.; Mason,~S.~A.; Albinati,~A.; Kozelka,~J.
  {Dispersion-Driven} Hydrogen Bonding: Predicted Hydrogen Bond between Water
  and {Platinum(II)} Identified by Neutron Diffraction. \emph{Angew. Chem. Int.
  Ed.} \textbf{2010}, \emph{49}, 7440--7443\relax
\mciteBstWouldAddEndPuncttrue
\mciteSetBstMidEndSepPunct{\mcitedefaultmidpunct}
{\mcitedefaultendpunct}{\mcitedefaultseppunct}\relax
\EndOfBibitem
\bibitem[Kozelka \latin{et~al.}(2000)Kozelka, Berg{\`e}s, Attias, and
  Fraitag]{kozelka_angew_2000}
Kozelka,~J.; Berg{\`e}s,~J.; Attias,~R.; Fraitag,~J. Hydrogen Bond with a
  Strong Dispersion Component. \emph{Angew. Chem. Int. Ed.} \textbf{2000},
  \emph{39}, 198--201\relax
\mciteBstWouldAddEndPuncttrue
\mciteSetBstMidEndSepPunct{\mcitedefaultmidpunct}
{\mcitedefaultendpunct}{\mcitedefaultseppunct}\relax
\EndOfBibitem
\bibitem[Lopes \latin{et~al.}(2008)Lopes, Rocha, Dos~Santos, and
  De~Almeida]{lopes_theoretical_2008}
Lopes,~J.~F.; Rocha,~W.~R.; Dos~Santos,~H.~F.; De~Almeida,~W.~B. Theoretical
  study of the potential energy surface for the interaction of cisplatin and
  their aquated species with water. \emph{J. Chem. Phys.} \textbf{2008},
  \emph{128}, 165103--14\relax
\mciteBstWouldAddEndPuncttrue
\mciteSetBstMidEndSepPunct{\mcitedefaultmidpunct}
{\mcitedefaultendpunct}{\mcitedefaultseppunct}\relax
\EndOfBibitem
\bibitem[Aono \latin{et~al.}(2016)Aono, Mori, and Sakaki]{aono_jctc_2016}
Aono,~S.; Mori,~T.; Sakaki,~S. 3D-{RISM}-{MP}2 {Approach} to {Hydration}
  {Structure} of {Pt}({II}) and {Pd}({II}) {Complexes}: {Unusual} {H}-{Ahead}
  {Mode} vs {Usual} {O}-{Ahead} {One}. \emph{J. Chem. Theory Comput.}
  \textbf{2016}, \emph{12}, 1189--1206\relax
\mciteBstWouldAddEndPuncttrue
\mciteSetBstMidEndSepPunct{\mcitedefaultmidpunct}
{\mcitedefaultendpunct}{\mcitedefaultseppunct}\relax
\EndOfBibitem
\bibitem[Berg{\`e}s \latin{et~al.}(2013)Berg{\`e}s, Fourr{\'e}, Pilm{\'e}, and
  Kozelka]{berges_inorgchem_2013}
Berg{\`e}s,~J.; Fourr{\'e},~I.; Pilm{\'e},~J.; Kozelka,~J. Quantum {Chemical}
  {Topology} {Study} of the {Water}-{Platinum}({II}) {Interaction}.
  \emph{Inorg. Chem.} \textbf{2013}, \emph{52}, 1217--1227\relax
\mciteBstWouldAddEndPuncttrue
\mciteSetBstMidEndSepPunct{\mcitedefaultmidpunct}
{\mcitedefaultendpunct}{\mcitedefaultseppunct}\relax
\EndOfBibitem
\bibitem[Melchior \latin{et~al.}(2015)Melchior, Tolazzi, Mart{\'i}nez,
  Pappalardo, and S{\'a}nchez~Marcos]{melchior_jctc_2015}
Melchior,~A.; Tolazzi,~M.; Mart{\'i}nez,~J.~M.; Pappalardo,~R.~R.;
  S{\'a}nchez~Marcos,~E. Hydration of {Two} {Cisplatin} {Aqua}-{Derivatives}
  {Studied} by {Quantum} {Mechanics} and {Molecular} {Dynamics} {Simulations}.
  \emph{J. Chem. Theory Comput.} \textbf{2015}, \emph{11}, 1735--1744\relax
\mciteBstWouldAddEndPuncttrue
\mciteSetBstMidEndSepPunct{\mcitedefaultmidpunct}
{\mcitedefaultendpunct}{\mcitedefaultseppunct}\relax
\EndOfBibitem
\bibitem[Kroutil \latin{et~al.}(2016)Kroutil, P{\v r}edota, and
  Chval]{kroutil_inorgchem_2016}
Kroutil,~O.; P{\v r}edota,~M.; Chval,~Z.
  Pt{\textperiodcentered}{\textperiodcentered}{\textperiodcentered}{H}
  {Nonclassical} {Interaction} in {Water}-{Dissolved} {Pt}({II}) {Complexes}:
  {Coaction} of {Electronic} {Effects} with {Solvent}-{Assisted}
  {Stabilization}. \emph{Inorg. Chem.} \textbf{2016}, \emph{55},
  3252--3264\relax
\mciteBstWouldAddEndPuncttrue
\mciteSetBstMidEndSepPunct{\mcitedefaultmidpunct}
{\mcitedefaultendpunct}{\mcitedefaultseppunct}\relax
\EndOfBibitem
\bibitem[Bancroft \latin{et~al.}(1990)Bancroft, Lepre, and
  Lippard]{bancroft_jacs_1990}
Bancroft,~D.~P.; Lepre,~C.~A.; Lippard,~S.~J. Platinum-195 {NMR} kinetic and
  mechanistic studies of cis- and trans-diamminedichloroplatinum({II}) binding
  to {DNA}. \emph{J. Am. Chem. Soc.} \textbf{1990}, \emph{112},
  6860--6871\relax
\mciteBstWouldAddEndPuncttrue
\mciteSetBstMidEndSepPunct{\mcitedefaultmidpunct}
{\mcitedefaultendpunct}{\mcitedefaultseppunct}\relax
\EndOfBibitem
\bibitem[Knox \latin{et~al.}(1986)Knox, Friedlos, Lydall, and
  Roberts]{knox_mechanism_1986}
Knox,~R.~J.; Friedlos,~F.; Lydall,~D.~A.; Roberts,~J.~J. Mechanism of
  {Cytotoxicity} of {Anticancer} {Platinum} {Drugs}: {Evidence} {That}
  cis-{Diamminedichloroplatinum}({II}) and
  cis-{Diammine}-(1,1-cyclobutanedicarboxylato)platinum({II}) {Differ} {Only}
  in the {Kinetics} of {Their} {Interaction} with {DNA}. \emph{Cancer Res.}
  \textbf{1986}, \emph{46}, 1972--1979\relax
\mciteBstWouldAddEndPuncttrue
\mciteSetBstMidEndSepPunct{\mcitedefaultmidpunct}
{\mcitedefaultendpunct}{\mcitedefaultseppunct}\relax
\EndOfBibitem
\bibitem[Alberts and Dorr(1998)Alberts, and Dorr]{alberts_new_1998}
Alberts,~D.~S.; Dorr,~R.~T. New {Perspectives} on an {Old} {Friend}:
  {Optimizing} {Carboplatin} for the {Treatment} of {Solid} {Tumors}. \emph{The
  Oncologist} \textbf{1998}, \emph{3}, 15--34\relax
\mciteBstWouldAddEndPuncttrue
\mciteSetBstMidEndSepPunct{\mcitedefaultmidpunct}
{\mcitedefaultendpunct}{\mcitedefaultseppunct}\relax
\EndOfBibitem
\bibitem[Johnstone \latin{et~al.}(2014)Johnstone, Alexander, Wilson, and
  Lippard]{johnstone_oxidative_2014}
Johnstone,~T.~C.; Alexander,~S.~M.; Wilson,~J.~J.; Lippard,~S.~J. Oxidative
  halogenation of cisplatin and carboplatin: synthesis, spectroscopy, and
  crystal and molecular structures of {Pt}({IV}) prodrugs. \emph{Dalton Trans.}
  \textbf{2014}, \emph{44}, 119--129\relax
\mciteBstWouldAddEndPuncttrue
\mciteSetBstMidEndSepPunct{\mcitedefaultmidpunct}
{\mcitedefaultendpunct}{\mcitedefaultseppunct}\relax
\EndOfBibitem
\bibitem[Blommaert \latin{et~al.}(1995)Blommaert, van Dijk-Knijnenburg, Dijt,
  den Engelse, Baan, Berends, and Fichtinger-Schepman]{blommaert_jacs_1995}
Blommaert,~F.~A.; van Dijk-Knijnenburg,~H. C.~M.; Dijt,~F.~J.; den Engelse,~L.;
  Baan,~R.~A.; Berends,~F.; Fichtinger-Schepman,~A. M.~J. Formation of {DNA}
  {Adducts} by the {Anticancer} {Drug} {Carboplatin}: {Different} {Nucleotide}
  {Sequence} {Preferences} in {Vitro} and in {Cells}. \emph{Biochemistry}
  \textbf{1995}, \emph{34}, 8474--8480\relax
\mciteBstWouldAddEndPuncttrue
\mciteSetBstMidEndSepPunct{\mcitedefaultmidpunct}
{\mcitedefaultendpunct}{\mcitedefaultseppunct}\relax
\EndOfBibitem
\bibitem[Zhang \latin{et~al.}(2001)Zhang, Guo, and You]{zhang_hydrolysis_2001}
Zhang,~Y.; Guo,~Z.; You,~X. Hydrolysis Theory for Cisplatin and Its Analogues
  Based on Density Functional Studies. \emph{J. Am. Chem. Soc.} \textbf{2001},
  \emph{123}, 9378--9387\relax
\mciteBstWouldAddEndPuncttrue
\mciteSetBstMidEndSepPunct{\mcitedefaultmidpunct}
{\mcitedefaultendpunct}{\mcitedefaultseppunct}\relax
\EndOfBibitem
\bibitem[Robertazzi and Platts(2004)Robertazzi, and
  Platts]{robertazzi_hydrogen_2004}
Robertazzi,~A.; Platts,~J.~A. Hydrogen bonding, solvation, and hydrolysis of
  cisplatin: A theoretical study. \emph{J. Comput. Chem.} \textbf{2004},
  \emph{25}, 1060--1067\relax
\mciteBstWouldAddEndPuncttrue
\mciteSetBstMidEndSepPunct{\mcitedefaultmidpunct}
{\mcitedefaultendpunct}{\mcitedefaultseppunct}\relax
\EndOfBibitem
\bibitem[Lau and Deubel(2006)Lau, and Deubel]{lau_hydrolysis_2006}
Lau,~J.~K.; Deubel,~D.~V. Hydrolysis of the Anticancer Drug Cisplatin:
  Pitfalls in the Interpretation of Quantum Chemical Calculations. \emph{J.
  Chem. Theory Comput.} \textbf{2006}, \emph{2}, 103--106\relax
\mciteBstWouldAddEndPuncttrue
\mciteSetBstMidEndSepPunct{\mcitedefaultmidpunct}
{\mcitedefaultendpunct}{\mcitedefaultseppunct}\relax
\EndOfBibitem
\bibitem[Alberto \latin{et~al.}(2009)Alberto, Lucas, Pavelka, and
  Russo]{alberto_second-generation_2009}
Alberto,~M.~E.; Lucas,~M. F.~A.; Pavelka,~M.; Russo,~N. The {Second-Generation}
  Anticancer Drug Nedaplatin: A Theoretical Investigation on the Hydrolysis
  Mechanism. \emph{J. Phys. Chem. B} \textbf{2009}, \emph{113},
  14473--14479\relax
\mciteBstWouldAddEndPuncttrue
\mciteSetBstMidEndSepPunct{\mcitedefaultmidpunct}
{\mcitedefaultendpunct}{\mcitedefaultseppunct}\relax
\EndOfBibitem
\bibitem[Lucas \latin{et~al.}(2009)Lucas, Pavelka, Alberto, and
  Russo]{lucas_neutral_2009}
Lucas,~M. F.~A.; Pavelka,~M.; Alberto,~M.~E.; Russo,~N. Neutral and Acidic
  Hydrolysis Reactions of the Third Generation Anticancer Drug Oxaliplatin.
  \emph{J. Phys. Chem. B} \textbf{2009}, \emph{113}, 831--838\relax
\mciteBstWouldAddEndPuncttrue
\mciteSetBstMidEndSepPunct{\mcitedefaultmidpunct}
{\mcitedefaultendpunct}{\mcitedefaultseppunct}\relax
\EndOfBibitem
\bibitem[Banerjee \latin{et~al.}(2010)Banerjee, Sengupta, and
  Mukherjee]{banerjee_cpl_2010}
Banerjee,~S.; Sengupta,~P.~S.; Mukherjee,~A.~K. Trans {Platinum} {Anticancer}
  {Drug} {AMD}443: {A} {Detailed} {Theoretical} {Study} by {DFT}-{TST} {Method}
  on the {Hydrolysis} {Mechanism}. \emph{Chem. Phys. Lett.} \textbf{2010},
  \emph{487}, 108--115\relax
\mciteBstWouldAddEndPuncttrue
\mciteSetBstMidEndSepPunct{\mcitedefaultmidpunct}
{\mcitedefaultendpunct}{\mcitedefaultseppunct}\relax
\EndOfBibitem
\bibitem[Melchior \latin{et~al.}(2010)Melchior, Marcos, Pappalardo, and
  Mart{\'i}nez]{melchior_comparative_2010}
Melchior,~A.; Marcos,~E.~S.; Pappalardo,~R.~R.; Mart{\'i}nez,~J.~M. Comparative
  study of the hydrolysis of a third- and a first-generation platinum
  anticancer complexes. \emph{Theor. Chem. Acc.} \textbf{2010}, \emph{128},
  627--638\relax
\mciteBstWouldAddEndPuncttrue
\mciteSetBstMidEndSepPunct{\mcitedefaultmidpunct}
{\mcitedefaultendpunct}{\mcitedefaultseppunct}\relax
\EndOfBibitem
\bibitem[Carloni \latin{et~al.}(2000)Carloni, Sprik, and
  Andreoni]{carloni_key_2000}
Carloni,~P.; Sprik,~M.; Andreoni,~W. Key Steps of the {cis-Platin-DNA}
  Interaction: Density Functional {Theory-Based} Molecular Dynamics
  Simulations. \emph{J. Phys. Chem. B} \textbf{2000}, \emph{104},
  823--835\relax
\mciteBstWouldAddEndPuncttrue
\mciteSetBstMidEndSepPunct{\mcitedefaultmidpunct}
{\mcitedefaultendpunct}{\mcitedefaultseppunct}\relax
\EndOfBibitem
\bibitem[Lau and Ensing(2010)Lau, and Ensing]{lau_pccp_2010}
Lau,~J. K.-C.; Ensing,~B. Hydrolysis of cisplatin{\textemdash}a
  first-principles metadynamics study. \emph{Phys. Chem. Chem. Phys.}
  \textbf{2010}, \emph{12}, 10348--10355\relax
\mciteBstWouldAddEndPuncttrue
\mciteSetBstMidEndSepPunct{\mcitedefaultmidpunct}
{\mcitedefaultendpunct}{\mcitedefaultseppunct}\relax
\EndOfBibitem
\bibitem[Yokogawa \latin{et~al.}(2007)Yokogawa, Sato, and
  Sakaki]{yokogawa_jcp_2007}
Yokogawa,~D.; Sato,~H.; Sakaki,~S. New generation of the reference interaction
  site model self-consistent field method: {Introduction} of spatial electron
  density distribution to the solvation theory. \emph{J. Chem. Phys.}
  \textbf{2007}, \emph{126}, 244504\relax
\mciteBstWouldAddEndPuncttrue
\mciteSetBstMidEndSepPunct{\mcitedefaultmidpunct}
{\mcitedefaultendpunct}{\mcitedefaultseppunct}\relax
\EndOfBibitem
\bibitem[Yokogawa \latin{et~al.}(2009)Yokogawa, Sato, and
  Sakaki]{yokogawa_jcp_2009}
Yokogawa,~D.; Sato,~H.; Sakaki,~S. Analytical energy gradient for reference
  interaction site model self-consistent field explicitly including spatial
  electron density distribution. \emph{J. Chem. Phys.} \textbf{2009},
  \emph{131}, 214504\relax
\mciteBstWouldAddEndPuncttrue
\mciteSetBstMidEndSepPunct{\mcitedefaultmidpunct}
{\mcitedefaultendpunct}{\mcitedefaultseppunct}\relax
\EndOfBibitem
\bibitem[Yokogawa \latin{et~al.}(2011)Yokogawa, Ono, Sato, and
  Sakaki]{yokogawa_dalton_2011}
Yokogawa,~D.; Ono,~K.; Sato,~H.; Sakaki,~S. Theoretical study on aquation
  reaction of cis-platin complex: {RISM}{\textendash}{SCF}{\textendash}{SEDD},
  a hybrid approach of accurate quantum chemical method and statistical
  mechanics. \emph{Dalton Trans.} \textbf{2011}, \emph{40}, 11125--11130\relax
\mciteBstWouldAddEndPuncttrue
\mciteSetBstMidEndSepPunct{\mcitedefaultmidpunct}
{\mcitedefaultendpunct}{\mcitedefaultseppunct}\relax
\EndOfBibitem
\bibitem[Hollis \latin{et~al.}(1989)Hollis, Amundsen, and
  Stern]{hollis_jmedchem_1989}
Hollis,~L.~S.; Amundsen,~A.~R.; Stern,~E.~W. Chemical and biological properties
  of a new series of cis-diammineplatinum({II}) antitumor agents containing
  three nitrogen donors: cis-[{Pt}({NH}$_3$)$_2$({N}-donor) {Cl}]$^+$. \emph{J. Med.
  Chem.} \textbf{1989}, \emph{32}, 128--136\relax
\mciteBstWouldAddEndPuncttrue
\mciteSetBstMidEndSepPunct{\mcitedefaultmidpunct}
{\mcitedefaultendpunct}{\mcitedefaultseppunct}\relax
\EndOfBibitem
\bibitem[Brabec(2002)]{brabec_dna_2002}
Brabec,~V. {DNA} {Modifications} by antitumor platinum and ruthenium compounds:
  {Their} recognition and repair. \emph{Prog. Nucleic Acid Res. Mol. Biol.}
  \textbf{2002}, \emph{71}, 1--68\relax
\mciteBstWouldAddEndPuncttrue
\mciteSetBstMidEndSepPunct{\mcitedefaultmidpunct}
{\mcitedefaultendpunct}{\mcitedefaultseppunct}\relax
\EndOfBibitem
\bibitem[Bursova \latin{et~al.}(2005)Bursova, Kasparkova, Hofr, and
  Brabec]{bursova_biophys_2005}
Bursova,~V.; Kasparkova,~J.; Hofr,~C.; Brabec,~V. Effects of {Monofunctional}
  {Adducts} of {Platinum}({II}) {Complexes} on {Thermodynamic} {Stability} and
  {Energetics} of {DNA} {Duplexes}. \emph{Biophys. J.} \textbf{2005},
  \emph{88}, 1207--1214\relax
\mciteBstWouldAddEndPuncttrue
\mciteSetBstMidEndSepPunct{\mcitedefaultmidpunct}
{\mcitedefaultendpunct}{\mcitedefaultseppunct}\relax
\EndOfBibitem
\bibitem[Lovejoy \latin{et~al.}(2008)Lovejoy, Todd, Zhang, McCormick, D'Aquino,
  Reardon, Sancar, Giacomini, and Lippard]{lovejoy_pnas_2008}
Lovejoy,~K.~S.; Todd,~R.~C.; Zhang,~S.; McCormick,~M.~S.; D'Aquino,~J.~A.;
  Reardon,~J.~T.; Sancar,~A.; Giacomini,~K.~M.; Lippard,~S.~J.
  cis-{Diammine}(pyridine)chloroplatinum({II}), a monofunctional platinum({II})
  antitumor agent: {Uptake}, structure, function, and prospects. \emph{PNAS}
  \textbf{2008}, \emph{105}, 8902--8907\relax
\mciteBstWouldAddEndPuncttrue
\mciteSetBstMidEndSepPunct{\mcitedefaultmidpunct}
{\mcitedefaultendpunct}{\mcitedefaultseppunct}\relax
\EndOfBibitem
\bibitem[Wang \latin{et~al.}(2010)Wang, Zhu, Huang, and
  Lippard]{wang_pnas_2010}
Wang,~D.; Zhu,~G.; Huang,~X.; Lippard,~S.~J. X-ray structure and mechanism of
  {RNA} polymerase {II} stalled at an antineoplastic monofunctional
  platinum-{DNA} adduct. \emph{PNAS} \textbf{2010}, \emph{107},
  9584--9589\relax
\mciteBstWouldAddEndPuncttrue
\mciteSetBstMidEndSepPunct{\mcitedefaultmidpunct}
{\mcitedefaultendpunct}{\mcitedefaultseppunct}\relax
\EndOfBibitem
\bibitem[Zhu \latin{et~al.}(2012)Zhu, Myint, Ang, Song, and
  Lippard]{zhu_cancer_2012}
Zhu,~G.; Myint,~M.; Ang,~W.~H.; Song,~L.; Lippard,~S.~J. Monofunctional
  {Platinum}{\textendash}{DNA} {Adducts} {Are} {Strong} {Inhibitors} of
  {Transcription} and {Substrates} for {Nucleotide} {Excision} {Repair} in
  {Live} {Mammalian} {Cells}. \emph{Cancer Res.} \textbf{2012}, \emph{72},
  790--800\relax
\mciteBstWouldAddEndPuncttrue
\mciteSetBstMidEndSepPunct{\mcitedefaultmidpunct}
{\mcitedefaultendpunct}{\mcitedefaultseppunct}\relax
\EndOfBibitem
\bibitem[Park \latin{et~al.}(2012)Park, Wilson, Song, and
  Lippard]{park_pnas_2012}
Park,~G.~Y.; Wilson,~J.~J.; Song,~Y.; Lippard,~S.~J. Phenanthriplatin, a
  monofunctional {DNA}-binding platinum anticancer drug candidate with unusual
  potency and cellular activity profile. \emph{PNAS} \textbf{2012}, \emph{109},
  11987--11992\relax
\mciteBstWouldAddEndPuncttrue
\mciteSetBstMidEndSepPunct{\mcitedefaultmidpunct}
{\mcitedefaultendpunct}{\mcitedefaultseppunct}\relax
\EndOfBibitem
\bibitem[Johnstone \latin{et~al.}(2014)Johnstone, Alexander, Lin, and
  Lippard]{johnstone_jacs_2014}
Johnstone,~T.~C.; Alexander,~S.~M.; Lin,~W.; Lippard,~S.~J. Effects of
  {Monofunctional} {Platinum} {Agents} on {Bacterial} {Growth}:
  {A}~{Retrospective} {Study}. \emph{J. Am. Chem. Soc.} \textbf{2014},
  \emph{136}, 116--118\relax
\mciteBstWouldAddEndPuncttrue
\mciteSetBstMidEndSepPunct{\mcitedefaultmidpunct}
{\mcitedefaultendpunct}{\mcitedefaultseppunct}\relax
\EndOfBibitem
\bibitem[Bowler \latin{et~al.}(2006)Bowler, Choudhury, Gillan, and
  Miyazaki]{bowler_pssb_2006}
Bowler,~D.~R.; Choudhury,~R.; Gillan,~M.~J.; Miyazaki,~T. Recent progress with
  large-scale ab initio calculations: the {CONQUEST} code. \emph{phys. stat.
  sol. (b)} \textbf{2006}, \emph{243}, 989--1000\relax
\mciteBstWouldAddEndPuncttrue
\mciteSetBstMidEndSepPunct{\mcitedefaultmidpunct}
{\mcitedefaultendpunct}{\mcitedefaultseppunct}\relax
\EndOfBibitem
\bibitem[Bowler and Miyazaki(2010)Bowler, and Miyazaki]{bowler_jpcms_2010}
Bowler,~D.~R.; Miyazaki,~T. Calculations for millions of atoms with density
  functional theory: linear scaling shows its potential. \emph{J. Phys.:
  Condens. Matter} \textbf{2010}, \emph{22}, 074207\relax
\mciteBstWouldAddEndPuncttrue
\mciteSetBstMidEndSepPunct{\mcitedefaultmidpunct}
{\mcitedefaultendpunct}{\mcitedefaultseppunct}\relax
\EndOfBibitem
\bibitem[Hern{\'a}ndez \latin{et~al.}(1996)Hern{\'a}ndez, Gillan, and
  Goringe]{hernandez_prb_1996}
Hern{\'a}ndez,~E.; Gillan,~M.~J.; Goringe,~C.~M. Linear-scaling
  density-functional-theory technique: {The} density-matrix approach.
  \emph{Phys. Rev. B} \textbf{1996}, \emph{53}, 7147\relax
\mciteBstWouldAddEndPuncttrue
\mciteSetBstMidEndSepPunct{\mcitedefaultmidpunct}
{\mcitedefaultendpunct}{\mcitedefaultseppunct}\relax
\EndOfBibitem
\bibitem[Sankey and Niklewski(1989)Sankey, and Niklewski]{sankey_pao_1989}
Sankey,~O.~F.; Niklewski,~D.~J. Ab initio multicenter tight-binding model for
  molecular-dynamics simulations and other applications in covalent systems.
  \emph{Phys. Rev. B} \textbf{1989}, \emph{40}, 3979--3995\relax
\mciteBstWouldAddEndPuncttrue
\mciteSetBstMidEndSepPunct{\mcitedefaultmidpunct}
{\mcitedefaultendpunct}{\mcitedefaultseppunct}\relax
\EndOfBibitem
\bibitem[Junquera \latin{et~al.}(2001)Junquera, Paz, S{\'a}nchez-Portal, and
  Artacho]{junquera_pao_2001}
Junquera,~J.; Paz,~{\'O}.; S{\'a}nchez-Portal,~D.; Artacho,~E. Numerical atomic
  orbitals for linear-scaling calculations. \emph{Phys. Rev. B} \textbf{2001},
  \emph{64}, 235111\relax
\mciteBstWouldAddEndPuncttrue
\mciteSetBstMidEndSepPunct{\mcitedefaultmidpunct}
{\mcitedefaultendpunct}{\mcitedefaultseppunct}\relax
\EndOfBibitem
\bibitem[Torralba \latin{et~al.}(2008)Torralba, Todorovi{\'c},
  Br{\'a}zdov{\'a}, Choudhury, Miyazaki, Gillan, and Bowler]{torralba_pao_2008}
Torralba,~A.~S.; Todorovi{\'c},~M.; Br{\'a}zdov{\'a},~V.; Choudhury,~R.;
  Miyazaki,~T.; Gillan,~M.~J.; Bowler,~D.~R. Pseudo-atomic orbitals as basis
  sets for the {O}({N}) {DFT} code {CONQUEST}. \emph{J. Phys.: Condens.
  Matter} \textbf{2008}, \emph{20}, 294206\relax
\mciteBstWouldAddEndPuncttrue
\mciteSetBstMidEndSepPunct{\mcitedefaultmidpunct}
{\mcitedefaultendpunct}{\mcitedefaultseppunct}\relax
\EndOfBibitem
\bibitem[Troullier and Martins(1991)Troullier, and
  Martins]{troullier_ncpp_1991}
Troullier,~N.; Martins,~J.~L. Efficient pseudopotentials for plane-wave
  calculations. \emph{Phys. Rev. B} \textbf{1991}, \emph{43}, 1993--2006\relax
\mciteBstWouldAddEndPuncttrue
\mciteSetBstMidEndSepPunct{\mcitedefaultmidpunct}
{\mcitedefaultendpunct}{\mcitedefaultseppunct}\relax
\EndOfBibitem
\bibitem[Perdew \latin{et~al.}(1996)Perdew, Burke, and
  Ernzerhof]{perdew_generalized_1996}
Perdew,~J.~P.; Burke,~K.; Ernzerhof,~M. Generalized Gradient Approximation Made
  Simple. \emph{Phys. Rev. Lett.} \textbf{1996}, \emph{77}, 3865--3868\relax
\mciteBstWouldAddEndPuncttrue
\mciteSetBstMidEndSepPunct{\mcitedefaultmidpunct}
{\mcitedefaultendpunct}{\mcitedefaultseppunct}\relax
\EndOfBibitem
\bibitem[Lin \latin{et~al.}(2012)Lin, Seitsonen, Tavernelli, and
  Rothlisberger]{lin_JCTC_2012}
Lin,~I.-C.; Seitsonen,~A.~P.; Tavernelli,~I.; Rothlisberger,~U. Structure and
  {Dynamics} of {Liquid} {Water} from ab {Initio} {Molecular}
  {Dynamics}{\textemdash}{Comparison} of {BLYP}, {PBE}, and {revPBE} {Density}
  {Functionals} with and without van der {Waals} {Corrections}. \emph{J. Chem.
  Theory Comput.} \textbf{2012}, \emph{8}, 3902--3910\relax
\mciteBstWouldAddEndPuncttrue
\mciteSetBstMidEndSepPunct{\mcitedefaultmidpunct}
{\mcitedefaultendpunct}{\mcitedefaultseppunct}\relax
\EndOfBibitem
\bibitem[DiStasio \latin{et~al.}(2014)DiStasio, Santra, Li, Wu, and
  Car]{distasio_JCP_2014}
DiStasio,~R.~A.; Santra,~B.; Li,~Z.; Wu,~X.; Car,~R. The individual and
  collective effects of exact exchange and dispersion interactions on the ab
  initio structure of liquid water. \emph{J. Chem. Phys.} \textbf{2014},
  \emph{141}, 084502\relax
\mciteBstWouldAddEndPuncttrue
\mciteSetBstMidEndSepPunct{\mcitedefaultmidpunct}
{\mcitedefaultendpunct}{\mcitedefaultseppunct}\relax
\EndOfBibitem
\bibitem[Truflandier and Autschbach(2010)Truflandier, and
  Autschbach]{truflandier_probing_2010}
Truflandier,~L.~A.; Autschbach,~J. Probing the Solvent Shell with {195Pt}
  Chemical Shifts: Density Functional Theory Molecular Dynamics Study of {Pt(II)}
  and {Pt(IV)} Anionic Complexes in Aqueous Solution. \emph{J. Am. Chem. Soc.}
  \textbf{2010}, \emph{132}, 3472--3483\relax
\mciteBstWouldAddEndPuncttrue
\mciteSetBstMidEndSepPunct{\mcitedefaultmidpunct}
{\mcitedefaultendpunct}{\mcitedefaultseppunct}\relax
\EndOfBibitem
\bibitem[Verlet(1967)]{verlet_computer_1967}
Verlet,~L. Computer {"Experiments"} on Classical Fluids. I. Thermodynamical
  Properties of {Lennard-Jones} Molecules. \emph{Phys. Rev.} \textbf{1967},
  \emph{159}, 98--103\relax
\mciteBstWouldAddEndPuncttrue
\mciteSetBstMidEndSepPunct{\mcitedefaultmidpunct}
{\mcitedefaultendpunct}{\mcitedefaultseppunct}\relax
\EndOfBibitem
\bibitem[Martyna \latin{et~al.}(1992)Martyna, Klein, and
  Tuckerman]{martyna_nosehoover_1992}
Martyna,~G.~J.; Klein,~M.~L.; Tuckerman,~M. Nos{\'e}{\textendash}{Hoover}
  chains: {The} canonical ensemble via continuous dynamics. \emph{J. Chem.
  Phys.} \textbf{1992}, \emph{97}, 2635--2643\relax
\mciteBstWouldAddEndPuncttrue
\mciteSetBstMidEndSepPunct{\mcitedefaultmidpunct}
{\mcitedefaultendpunct}{\mcitedefaultseppunct}\relax
\EndOfBibitem
\bibitem[Hirakawa \latin{et~al.}(2017)Hirakawa, Suzuki, Bowler, and
  Miyazaki]{hirakawa_nosehoover_2017}
Hirakawa,~T.; Suzuki,~T.; Bowler,~D.~R.; Miyazaki,~T. Canonical-ensemble
  extended {Lagrangian} {Born}{\textendash}{Oppenheimer} molecular dynamics for
  the linear scaling density functional theory. \emph{J. Phys.: Condens.
  Matter} \textbf{2017}, \emph{29}, 405901\relax
\mciteBstWouldAddEndPuncttrue
\mciteSetBstMidEndSepPunct{\mcitedefaultmidpunct}
{\mcitedefaultendpunct}{\mcitedefaultseppunct}\relax
\EndOfBibitem
\bibitem[Carter \latin{et~al.}(1989)Carter, Ciccotti, Hynes, and
  Kapral]{carter1989constrained}
Carter,~E.; Ciccotti,~G.; Hynes,~J.~T.; Kapral,~R. Constrained reaction
  coordinate dynamics for the simulation of rare events. \emph{Chem. Phys.
  Lett.} \textbf{1989}, \emph{156}, 472--477\relax
\mciteBstWouldAddEndPuncttrue
\mciteSetBstMidEndSepPunct{\mcitedefaultmidpunct}
{\mcitedefaultendpunct}{\mcitedefaultseppunct}\relax
\EndOfBibitem
\bibitem[Sprik and Ciccotti(1998)Sprik, and Ciccotti]{sprik1998free}
Sprik,~M.; Ciccotti,~G. Free energy from constrained molecular dynamics.
  \emph{J. Chem. Phys.} \textbf{1998}, \emph{109}, 7737--7744\relax
\mciteBstWouldAddEndPuncttrue
\mciteSetBstMidEndSepPunct{\mcitedefaultmidpunct}
{\mcitedefaultendpunct}{\mcitedefaultseppunct}\relax
\EndOfBibitem
\bibitem[Ryckaert \latin{et~al.}(1977)Ryckaert, Ciccotti, and
  Berendsen]{ryckaert1977numerical}
Ryckaert,~J.-P.; Ciccotti,~G.; Berendsen,~H.~J. Numerical integration of the
  cartesian equations of motion of a system with constraints: molecular
  dynamics of n-alkanes. \emph{J. Comput. Phys.} \textbf{1977}, \emph{23},
  327--341\relax
\mciteBstWouldAddEndPuncttrue
\mciteSetBstMidEndSepPunct{\mcitedefaultmidpunct}
{\mcitedefaultendpunct}{\mcitedefaultseppunct}\relax
\EndOfBibitem
\bibitem[Andersen(1983)]{andersen1983rattle}
Andersen,~H.~C. Rattle: A ``velocity'' version of the shake algorithm for
  molecular dynamics calculations. \emph{J. Comput. Phys.} \textbf{1983},
  \emph{52}, 24--34\relax
\mciteBstWouldAddEndPuncttrue
\mciteSetBstMidEndSepPunct{\mcitedefaultmidpunct}
{\mcitedefaultendpunct}{\mcitedefaultseppunct}\relax
\EndOfBibitem
\bibitem[Ciccotti and Ryckaert(1986)Ciccotti, and
  Ryckaert]{ciccotti1986molecular}
Ciccotti,~G.; Ryckaert,~J.-P. Molecular dynamics simulation of rigid molecules.
  \emph{Comp. Phys. Rep.} \textbf{1986}, \emph{4}, 346--392\relax
\mciteBstWouldAddEndPuncttrue
\mciteSetBstMidEndSepPunct{\mcitedefaultmidpunct}
{\mcitedefaultendpunct}{\mcitedefaultseppunct}\relax
\EndOfBibitem
\bibitem[Jacucci and Rahman(1984)Jacucci, and Rahman]{jacucci_comparing_1984}
Jacucci,~G.; Rahman,~A. Comparing the efficiency of {Metropolis} {Monte}
  {Carlo} and molecular-dynamics methods for configuration space sampling.
  \emph{Il Nuovo Cimento D} \textbf{1984}, \emph{4}, 341--356\relax
\mciteBstWouldAddEndPuncttrue
\mciteSetBstMidEndSepPunct{\mcitedefaultmidpunct}
{\mcitedefaultendpunct}{\mcitedefaultseppunct}\relax
\EndOfBibitem
\bibitem[Giannozzi and {et al.}(2009)Giannozzi, and {et al.}]{qe_short_2009}
Giannozzi,~P.; {et al.}, \emph{J. Phys.: Condens. Matter} \textbf{2009},
  \emph{21}, 395502\relax
\mciteBstWouldAddEndPuncttrue
\mciteSetBstMidEndSepPunct{\mcitedefaultmidpunct}
{\mcitedefaultendpunct}{\mcitedefaultseppunct}\relax
\EndOfBibitem
\bibitem[Frisch and {et al.}()Frisch, and {et al.}]{gaussian09_short}
Frisch,~M.~J.; {et al.}, Gaussian09. \emph{Gaussian 09, Revision A.02;
  \normalfont{Gaussian, Inc., Wallingford, CT, 2009}} \relax
\mciteBstWouldAddEndPunctfalse
\mciteSetBstMidEndSepPunct{\mcitedefaultmidpunct}
{}{\mcitedefaultseppunct}\relax
\EndOfBibitem
\bibitem[Press \latin{et~al.}(1992)Press, Flannery, Teukolsky, and
  Vetterling]{press_numerical_1992}
Press,~W.~H.; Flannery,~B.~P.; Teukolsky,~S.~A.; Vetterling,~W.~T.
  \emph{Numerical {Recipes} in {Fortran} 77: {The} {Art} of {Scientific}
  {Computing}}, 2nd ed.; Cambridge University Press: Cambridge England ; New
  York, NY, USA, 1992\relax
\mciteBstWouldAddEndPuncttrue
\mciteSetBstMidEndSepPunct{\mcitedefaultmidpunct}
{\mcitedefaultendpunct}{\mcitedefaultseppunct}\relax
\EndOfBibitem
\bibitem[Lau and Ensing(2010)Lau, and Ensing]{lau_hydrolysis_2010}
Lau,~J.~K.; Ensing,~B. Hydrolysis of cisplatin—a first-principles
  metadynamics study. \emph{Phys. Chem. Chem. Phys.} \textbf{2010}, \relax
\mciteBstWouldAddEndPunctfalse
\mciteSetBstMidEndSepPunct{\mcitedefaultmidpunct}
{}{\mcitedefaultseppunct}\relax
\EndOfBibitem
\bibitem[Nakamoto \latin{et~al.}(1965)Nakamoto, McCarthy, Fujita, Condrate, and
  Behnke]{nakamoto_infrared_1965}
Nakamoto,~K.; McCarthy,~P.~J.; Fujita,~J.; Condrate,~R.~A.; Behnke,~G.~T.
  Infrared {Studies} of {Ligand}-{Ligand} {Interaction} in
  {Dihalogenodiammineplatinum}({II}) {Complexes}. \emph{Inorg. Chem.}
  \textbf{1965}, \emph{4}, 36--43\relax
\mciteBstWouldAddEndPuncttrue
\mciteSetBstMidEndSepPunct{\mcitedefaultmidpunct}
{\mcitedefaultendpunct}{\mcitedefaultseppunct}\relax
\EndOfBibitem
\bibitem[Wysoki{\'n}ski and Michalska(2001)Wysoki{\'n}ski, and
  Michalska]{wysokinski_performance_2001}
Wysoki{\'n}ski,~R.; Michalska,~D. The performance of different density
  functional methods in the calculation of molecular structures and vibrational
  spectra of {platinum(II)} antitumor drugs: cisplatin and carboplatin.
  \emph{J. Comput. Chem.} \textbf{2001}, \emph{22}, 901--912\relax
\mciteBstWouldAddEndPuncttrue
\mciteSetBstMidEndSepPunct{\mcitedefaultmidpunct}
{\mcitedefaultendpunct}{\mcitedefaultseppunct}\relax
\EndOfBibitem
\bibitem[Raugei \latin{et~al.}(1999)Raugei, Cardini, and
  Schettino]{raugei_ab_1999}
Raugei,~S.; Cardini,~G.; Schettino,~V. An \textit{ab initio} molecular dynamics
  study of the {SN}$_2$ reaction {Cl}$^-$+{CH}$_3$Br{$\rightarrow$}{CH}$_3$Cl+{Br}$^-$.
  \emph{The Journal of Chemical Physics} \textbf{1999}, \emph{111},
  10887--10894\relax
\mciteBstWouldAddEndPuncttrue
\mciteSetBstMidEndSepPunct{\mcitedefaultmidpunct}
{\mcitedefaultendpunct}{\mcitedefaultseppunct}\relax
\EndOfBibitem
\bibitem[Yang \latin{et~al.}(2004)Yang, Fleurat-Lessard, Hristov, and
  Ziegler]{yang2004free}
Yang,~S.-Y.; Fleurat-Lessard,~P.; Hristov,~I.; Ziegler,~T. Free Energy Profiles
  for the Identity SN2 Reactions Cl$^-$ + CH$_3$Cl and NH$_3$ + H$_3$BNH$_3$: A Constraint Ab
  Initio Molecular Dynamics Study. \emph{J. Phys. Chem. A} \textbf{2004},
  \emph{108}, 9461--9468\relax
\mciteBstWouldAddEndPuncttrue
\mciteSetBstMidEndSepPunct{\mcitedefaultmidpunct}
{\mcitedefaultendpunct}{\mcitedefaultseppunct}\relax
\EndOfBibitem
\bibitem[B{\"u}hl and Kabrede(2006)B{\"u}hl, and Kabrede]{buhl_mechanism_2006}
B{\"u}hl,~M.; Kabrede,~H. Mechanism of {Water} {Exchange} in {Aqueous}
  {Uranyl}({VI}) {Ion}. {A} {Density} {Functional} {Molecular} {Dynamics}
  {Study}. \emph{Inorg. Chem.} \textbf{2006}, \emph{45}, 3834--3836\relax
\mciteBstWouldAddEndPuncttrue
\mciteSetBstMidEndSepPunct{\mcitedefaultmidpunct}
{\mcitedefaultendpunct}{\mcitedefaultseppunct}\relax
\EndOfBibitem
\bibitem[Komeiji \latin{et~al.}(2009)Komeiji, Ishikawa, Mochizuki, Yamataka,
  and Nakano]{komeiji_fragment_2009}
Komeiji,~Y.; Ishikawa,~T.; Mochizuki,~Y.; Yamataka,~H.; Nakano,~T. Fragment
  {Molecular} {Orbital} method-based {Molecular} {Dynamics} ({FMO}-{MD}) as a
  simulator for chemical reactions in explicit solvation. \emph{J. Comput.
  Chem.} \textbf{2009}, \emph{30}, 40--50\relax
\mciteBstWouldAddEndPuncttrue
\mciteSetBstMidEndSepPunct{\mcitedefaultmidpunct}
{\mcitedefaultendpunct}{\mcitedefaultseppunct}\relax
\EndOfBibitem
\bibitem[Miller and House(1989)Miller, and House]{miller_hydrolysis_1989_1}
Miller,~S.~E.; House,~D.~A. The hydrolysis products of
  cis-diamminedichloroplatinum({II}). {1}. {The} kinetics of formation and
  anation of the cis-diammine(aqua)chloroplatinum({II}) cation in acidic
  aqueous solution. \emph{Inorg. Chim. Acta} \textbf{1989}, \emph{161},
  131--137\relax
\mciteBstWouldAddEndPuncttrue
\mciteSetBstMidEndSepPunct{\mcitedefaultmidpunct}
{\mcitedefaultendpunct}{\mcitedefaultseppunct}\relax
\EndOfBibitem
\bibitem[Miller and House(1989)Miller, and House]{miller_hydrolysis_1989_2}
Miller,~S.~E.; House,~D.~A. The hydrolysis products of
  cis-dichlorodiammineplatinum({II}) 2. {The} kinetics of formation and anation
  of the cis-diamminedi(aqua)platinum({II}) cation. \emph{Inorg. Chim. Acta}
  \textbf{1989}, \emph{166}, 189--197\relax
\mciteBstWouldAddEndPuncttrue
\mciteSetBstMidEndSepPunct{\mcitedefaultmidpunct}
{\mcitedefaultendpunct}{\mcitedefaultseppunct}\relax
\EndOfBibitem
\bibitem[Miller and House(1990)Miller, and House]{miller_hydrolysis_1990}
Miller,~S.~E.; House,~D.~A. The hydrolysis products of
  cis-dichlorodiammineplatinum({II}) 3. {Hydrolysis} kinetics at physiological
  {pH}. \emph{Inorg. Chim. Acta} \textbf{1990}, \emph{173}, 53--60\relax
\mciteBstWouldAddEndPuncttrue
\mciteSetBstMidEndSepPunct{\mcitedefaultmidpunct}
{\mcitedefaultendpunct}{\mcitedefaultseppunct}\relax
\EndOfBibitem
\bibitem[Reishus and Martin(1961)Reishus, and Martin]{reishus_cisplatin_1961}
Reishus,~J.~W.; Martin,~D.~S. cis-{Dichlorodiammineplatinum}({II}). {Acid}
  {Hydrolysis} and {Isotopic} {Exchange} of the {Chloride} {Ligands}. \emph{J.
  Am. Chem. Soc.} \textbf{1961}, \emph{83}, 2457--2462\relax
\mciteBstWouldAddEndPuncttrue
\mciteSetBstMidEndSepPunct{\mcitedefaultmidpunct}
{\mcitedefaultendpunct}{\mcitedefaultseppunct}\relax
\EndOfBibitem
\bibitem[Marti \latin{et~al.}(1998)Marti, Bon~Hoa, and
  Kozelka]{marti_reversible_1998}
Marti,~N.; Bon~Hoa,~G.~H.; Kozelka,~J. Reversible hydrolysis of [{PtCl}(dien)]$^+$
  and [{PtCl}({NH}$_3$)$_3$]$^+$. {Determination} of the rate constants using {UV}
  spectrophotometry. \emph{Inorg. Chem. Comm.} \textbf{1998}, \emph{1},
  439--442\relax
\mciteBstWouldAddEndPuncttrue
\mciteSetBstMidEndSepPunct{\mcitedefaultmidpunct}
{\mcitedefaultendpunct}{\mcitedefaultseppunct}\relax
\EndOfBibitem
\bibitem[Davies \latin{et~al.}(2000)Davies, Berners-Price, and
  Hambley]{davies_slowing_2000}
Davies,~M.~S.; Berners-Price,~S.~J.; Hambley,~T.~W. Slowing of {Cisplatin}
  {Aquation} in the {Presence} of {DNA} but {Not} in the {Presence} of
  {Phosphate}: {Improved} {Understanding} of {Sequence} {Selectivity} and the
  {Roles} of {Monoaquated} and {Diaquated} {Species} in the {Binding} of
  {Cisplatin} to {DNA}. \emph{Inorg. Chem.} \textbf{2000}, \emph{39},
  5603--5613\relax
\mciteBstWouldAddEndPuncttrue
\mciteSetBstMidEndSepPunct{\mcitedefaultmidpunct}
{\mcitedefaultendpunct}{\mcitedefaultseppunct}\relax
\EndOfBibitem
\bibitem[Lee and Martin(1976)Lee, and Martin]{lee_cisplatin_1976}
Lee,~K.~W.; Martin,~D.~S. Cis-dichlorodiammineplatinum({II}). {Aquation}
  equilibria and isotopic exchange of chloride ligands with free chloride and
  tetrachloroplatinate({II}). \emph{Inorg. Chim. Acta} \textbf{1976},
  \emph{17}, 105--110\relax
\mciteBstWouldAddEndPuncttrue
\mciteSetBstMidEndSepPunct{\mcitedefaultmidpunct}
{\mcitedefaultendpunct}{\mcitedefaultseppunct}\relax
\EndOfBibitem
\bibitem[Vinje \latin{et~al.}(2005)Vinje, Sletten, and
  Kozelka]{vinje_influence_2005}
Vinje,~J.; Sletten,~E.; Kozelka,~J. Influence of {dT}$_{20}$ and [d({AT})$_{10}$]$_2$ on
  {Cisplatin} {Hydrolysis} {Studied} by {Two}-{Dimensional} [$^{1}$H,$^{15}$N] {HMQC}
  {NMR} {Spectroscopy}. \emph{Chem. Eur. J.} \textbf{2005}, \emph{11},
  3863--3871\relax
\mciteBstWouldAddEndPuncttrue
\mciteSetBstMidEndSepPunct{\mcitedefaultmidpunct}
{\mcitedefaultendpunct}{\mcitedefaultseppunct}\relax
\EndOfBibitem
\bibitem[Zhang \latin{et~al.}(2013)Zhang, Pham, Gygi, and
  Galli]{zhang_chlorine_2013}
Zhang,~C.; Pham,~T.~A.; Gygi,~F.; Galli,~G. Communication: {Electronic}
  structure of the solvated chloride anion from first principles molecular
  dynamics. \emph{J. Chem. Phys.} \textbf{2013}, \emph{138}, 181102\relax
\mciteBstWouldAddEndPuncttrue
\mciteSetBstMidEndSepPunct{\mcitedefaultmidpunct}
{\mcitedefaultendpunct}{\mcitedefaultseppunct}\relax
\EndOfBibitem
\end{mcitethebibliography}
\end{document}